\documentclass[a4paper,fleqn,usenatbib]{mnras}

\usepackage{txfonts}

\usepackage[T1]{fontenc}
\usepackage{ae,aecompl}

\usepackage{graphicx}	
\usepackage{xcolor}

\newcommand{\e}[1]{\times 10^{#1}}

\title[Entrainment in 1D stellar models]{Convective core entrainment in 1D main sequence stellar models}

\author[L. J. A. Scott et al.]{
L. J. A. Scott,$^{1}$\thanks{E-mail: l.j.a.scott@keele.ac.uk}
R. Hirschi,$^{1,2}$
C. Georgy,$^{3}$
W. D. Arnett,$^{4}$
C. Meakin,$^{4,5}$
\newauthor{E. A. Kaiser,$^{1}$
S. Ekstr{\"o}m$^{3}$
and N. Yusof}$^{6}$
\\
$^{1}$Astrophysics Group, Lennard-Jones Laboratories, Keele University, Keele ST5 5BG, UK\\
$^{2}$Institute for the Physics and Mathematics of the Universe (WPI), University of Tokyo, 5-1-5 Kashiwanoha, Kashiwa 277-8583, Japan\\
$^{3}$Department of Astronomy, University of Geneva, Ch. Maillettes 51, 1290 Versoix, Switzerland\\
$^{4}$Steward Observatory, University of Arizona, 933 N. Cherry Avenue, Tucson AZ 85721, USA\\
$^{5}$Pasadena Consulting Group, 1075 N Mar Vista Ave, Pasadena, CA 91104 USA\\
$^{6}$Department of Physics, Faculty of Science, University of Malaya, 50603 Kuala Lumpur, Malaysia\\
}
\date{Accepted XXX. Received YYY; in original form ZZZ}

\pubyear{2020}

\begin{document}
\label{firstpage}
\pagerange{\pageref{firstpage}--\pageref{lastpage}}
\maketitle

\begin{abstract}
3D hydrodynamics models of deep stellar convection exhibit turbulent entrainment at the convective-radiative boundary which follows the entrainment law, varying with boundary penetrability.
We implement the entrainment law in the 1D Geneva stellar evolution code. We then calculate models between 1.5 and 60\,M$_{\odot}$ at solar metallicity ($Z=0.014$) and compare them to previous generations of models and observations on the main sequence.
The boundary penetrability, quantified by the bulk Richardson number, $Ri_{\rm B}$, varies with mass and to a smaller extent with time. The variation of $Ri_{\rm B}$ with mass is due to the mass dependence of typical convective velocities in the core and hence the luminosity of the star. The chemical gradient above the convective core dominates the variation of $Ri_{\rm B}$ with time.
An entrainment law method can therefore explain the apparent mass dependence of convective boundary mixing through $Ri_{\rm B}$.
New models including entrainment can better reproduce the mass dependence of the main sequence width using entrainment law parameters $A\sim2\e{-4}$ and $n=1$.
We compare these empirically constrained values to the results of 3D hydrodynamics simulations and discuss implications.

\end{abstract}




\section{Introduction}

It has long been known that convective boundary mixing (CBM) must be included into stellar models in order to reproduce observations.
The main sequence (MS) width of clusters is one of the best-known examples of such observations;
other examples include large samples of wide binaries and asteroseismic measurements \citep[e.\,g.][]{claret2019,deheuvels2016}.
As a result, stellar models' CBM schemes are calibrated to give results consistent with the observed reality.
\citet{castro2014} showed that current generations of models have MS widths on the Hertzsprung-Russell diagram (HRD) which are too narrow for high mass stars.
The discrepancy in width grows larger with mass.

Currently, CBM is usually implemented in 1D stellar evolution codes in one of two ways. The first of these is step overshoot, which is an extension of the convective core by some fraction of a pressure scale height \citep[see e.\,g.][]{ekstrom2012}. Depending on the code, mixing in the overshoot region could be instantaneous or diffusive. The second is exponentially decaying diffusion, where mixing is governed by a diffusion coefficient which decays exponentially from a value near the Schwarzschild boundary \citep{freytag1996,herwig2000}. The parameters in both can be calibrated in order to match observations such as post-MS spin down \citep[Fig.\,1 in][]{brott2011} and asteroseismic frequencies \citep{aerts2018}.

Both 3D hydrodynamics simulations and observations can be compared to 1D models incorporating CBM and other mixing processes, such as waves \citep{meakin2007,jones2017,edelmann2019,muller2020,pratt2020}.
Incorporating 3D hydrodynamics results, such as convective regions which grow as a result of entrainment, into 1D models also allows them to be studied on evolutionary timescales.
It has been shown that the rate of entrainment of material at convective borders is dependent on the bulk Richardson number, $Ri_{\rm B}$, a dimensionless measure of the penetrability of a boundary by convection.
For example, \citet{cristini2019} showed that both the upper and lower boundaries in a convective shell followed the same entrainment law, suggesting that CBM is controlled by the global properties of the convective region.
Despite these results from simulation, the entrainment law is not widely used in 1D stellar evolution codes.

Prior to this study, only \citet{staritsin2013} has published 1D entrainment law stellar models. Staritsin's models of 16 and 24\,M$_\odot$ main sequence stars with entrainment were calibrated using asteroseismology values for the extent of mixing. In these models, the extent of extra mixing beyond the formally convective region (in units of pressure scale heights) decreased as the models evolved. This contrasts traditional CBM which typically stays constant.

In this paper, we investigate entrainment in 1D main sequence models from 1.5 to 60\,M$_{\odot}$ using the Geneva stellar evolution code.
We then compare our new models with entrainment to models including standard overshoot and constrain our entrainment parameters using these models.
We further constrain our entrainment models with comparison to observations, in particular the MS width.
We contrast entrainment parameters constrained by observations with those obtained in 3D simulations and discuss implications.
Section\,\ref{sec:methods} explains the definition and calculation of $Ri_{\rm B}$ and the entrainment algorithm, along with the parameters of the model grid. Section\,\ref{sec:results} discusses the properties of the models, focussing on $Ri_{\rm B}$ and the entrainment parameters. We present our conclusions in Section\,\ref{sec:discussion} and discuss the the plausibility and implications of our entrainment algorithm.

\section{Methods}
\label{sec:methods}
\subsection{Calculation of Bulk Richardson number}
\label{sec:calc_rib}
The bulk Richardson number is defined as
\begin{equation}
\label{eq:rib}
  Ri_{\rm B} = \frac{l_{\rm c} \Delta b}{v_{\rm c}^2},
\end{equation}
where $l_{\rm c}$ is a length-scale for turbulent motions in the convective region, $v_{\rm c}$ is the typical speed of convective flows and $\Delta b$ is the buoyancy jump. For $l_{\rm c}$ we use $0.5H_{P,\rm b}$, where $H_{P,\rm b}$ is the pressure scale height at $r=r_{\rm b}$ and $r_{\rm b}$ is the radius of the convective core.
If there is no CBM included in the model, $r_{\rm b}$ is equivalent to the Schwarzschild boundary.
Otherwise, it is the radius to which the CBM extends.

$l_{\rm c}$ represents the length-scale of the largest fluid elements in the turbulent region. The motivation for our choice of $l_{\rm c}=0.5H_{P.\rm b}$ comes from the results of \citet{meakin2007}, who found that the horizontal correlation length-scale for velocity in their simulation of convection was approximately half a pressure scale height. We aim to be consistent with \citet{cristini2019} by using this estimate of the horizontal correlation length-scale as a proxy for $l_c$.

The buoyancy jump is an integral of the squared buoyancy frequency $N^2$ with respect to radius $r$, given by
\begin{equation}
\label{eq:buoyjump}
  \Delta b = \int_{r_1}^{r_2}
             N^2 \mathrm{d}r,
\end{equation}
where $r_1$ and $r_2$ encompass the boundary of the convective core, centred at $r=r_{\rm b}$.
The upper limit $r_2$ is equal to $r_{\rm b}$ plus some fraction of a pressure scale height;
in this study we use $r_2=r_{\rm b}+0.25H_{P,\rm b}$ to be consistent with previous work \citep{cristini2019}.
Conversely, $r_1$ is the larger of either $r_{\rm b}-0.25 H_{P,\rm b}$ or the Schwarzschild boundary.
Using this maximum prevents negative $N^2$ regions, which do not contribute to buoyancy braking, from being included in the buoyancy jump integration.

In our prescription, the size of the integration region encompassed by $r_1$ and $r_2$ is between $0.25$ and $0.5 H_{P,\rm b}$, depending on the size of the entrainment region. This is supposed to encompass the part of the boundary in which fluid elements are decelerated and turned back towards the convective region by buoyancy. \citet{cristini2019} Table\,2 gives examples of their simulation boundary widths, which are all a fraction (0.1 to 0.6) of a pressure scale height. We cannot use a boundary width of $\sim 2v_{\rm c}/N$ as in \citet{staritsin2013}, since at our boundary we have $N=0$, so we use the approach of \citet{cristini2019}. However, we cannot be sure if the boundary width does not vary with mass (other than what is already contained in the mass dependence of $H_{P,\rm b}$), and the integration region size must still be considered a free parameter. The buoyancy jump and its dependence on these parameters is discussed in more detail in Section\,\ref{sec:time_dependence} and Fig.\,\ref{fig:N2profiles}.

For the buoyancy frequency, we use
\begin{equation}
 \label{eq:buoyfreq}
 N^2 = \frac{g\delta}{H_P}(\nabla_{\rm ad} - \nabla + \frac{1}{\delta}\nabla_{\rm \mu}),
\end{equation}
where $g$ is the gravitational acceleration, $H_P$ the pressure scale height, $\delta$ the density gradient with respect to temperature $(-\frac{\partial \ln \rho}{\partial \ln T})$, $\nabla_{\rm ad}$ the adiabatic temperature gradient, $\nabla$ the actual temperature gradient, and $\nabla_{\rm \mu}$ the mean molecular weight gradient.  


For $v_{\rm c}$, we use a mass-weighted root mean square of the mixing length theory (MLT) velocity, $v_{\rm{MLT}}$, in the core:

\begin{equation}
\label{eq:velocity}
v_{\mathrm{c}}= 
  \sqrt{\frac{\sum_i v_{\mathrm{MLT,}i}^2 \Delta m_i}
  {\sum_i \Delta m_i}},
\end{equation}
where $i$ represents the model mesh point and $\Delta m$ is the mass contained between the midpoints of shells $i+1$ and $i$, and $i-1$ and $i$. This sum over $i$ is taken from the centre of the core to the Schwarzschild boundary. This average value is less subject to fluctuation due to numerical factors such as zoning than a single value chosen at some distance from the boundary. $v_{\rm c}$ estimates the typical speed of the flow in the convective region which is responsible for entrainment.

\subsection{Entrainment law algorithm}
\label{sec:algorithm}
$Ri_{\mathrm{B}}$ is a good measure of how difficult it is for convective flows to entrain material from the stable region. Indeed, the numerator, $l_{\rm c}\Delta b$, measures the stability or stiffness of the convective boundary region via the buoyancy frequency $N^2$. The denominator, $v_{\mathrm{c}}^2$ ($\propto$ specific kinetic energy), measures the vigour of the convective flows approaching the boundary. A higher $Ri_{\mathrm{B}}$ value thus means that it is harder for convection to entrain material from the stable region above.

We then use $Ri_{\mathrm{B}}$ in the entrainment law\footnote{Whilst the use of the word 'law' suggests that all the parameters have a determined value, this is not the case for the entrainment law. However, since this is the currently accepted terminology in other fields such as geophysics, we will continue to use it.} \citep[e.\,g.][]{fernando1991} to calculate an entrainment rate.
The entrainment law is
\begin{equation}
\label{eq:entrainmentlaw}
  \frac{v_{\mathrm{e}}}{v_{\mathrm{c}}}=ARi_{\mathrm{B}}^{-n},
\end{equation}
where $v_{\mathrm{e}}$ is the convective boundary progression speed and $A$ and $n$ are parameters controlling the entrainment rate.
Note that if $n=1$ (as is the case for most of our models), any uncertainty in $l_{\rm c}$ in Eq.\,\ref{eq:rib} would inversely scale $A$. However, we are not targeting exact values for these parameters, and we can be fairly certain given the results of \citet{meakin2007} that $l_{\rm c}\sim H_{P,\rm b}$ as we have assumed.

A mass entrainment rate, $\dot{M}_{\rm{ent}}$, can be derived from this to give
\begin{equation}
\label{eq:mdot}
  \dot{M}_{\rm{ent}}=4\pi r_{\mathrm{b}}^2 \rho_{\mathrm{b}} v_{\mathrm{c}}ARi_{\mathrm{B}}^{-n},
\end{equation}
with $\rho_{\mathrm{b}}$ being the density at $r=r_{\mathrm{b}}$.
The mass contained within the entrained region, $M_{\mathrm{ent}}$, is then
\begin{equation}
\label{eq:ment}
  M_{\rm{ent}} = \sum_j \dot{M}_{\mathrm{ent},j} \mathrm{\Delta}t_j
\end{equation}
where $j$ denotes the model time step with length $\mathrm{\Delta}t$.
This region is then considered part of the convective core. This means that the region is then instantaneously mixed and the temperature gradient is set to $\nabla_{\rm ad}$ (further discussed in Section\,\ref{sec:discussion}).

In our implementation of the entrainment law, the entrained mass accumulates over the lifetime of the core with each time step, according to Eq.\,\ref{eq:ment}.
Since the value of $Ri_{\rm B}$ controls $\dot{M}_{\mathrm{ent},j}$ rather than $M_{\rm{ent}}$ directly, any previous history of entrainment in the models is unaffected by the instantaneous value of $Ri_{\rm B}$.
This contrasts the previous implementation of \citet{staritsin2013}, in which the entrained distance at any time step is equal to $v_{\rm e}\rm{\Delta}t$. Thus, our prescription can be viewed as cumulative entrainment and Staritsin's as instantaneous (meaning that it depends only on the stellar structure at the current time step).
3D hydrodynamic simulations of stellar convection which exhibit entrainment show that the convective region continuously accumulates material. This is the motivation for a cumulative entrainment method, as once material is entrained, it stays well-mixed. However, it is not known whether this holds true on evolutionary time-scales so it is not clear at this point which approach is more appropriate. Our method allows us to investigate the consequences of cumulative entrainment which is controlled by the changing value of the bulk Richardson number and we compare our results to \citet{staritsin2013} in Section\,\ref{sec:discussion}.

\subsection{Geneva code model grid}
\label{sec:grid}
We use the Geneva stellar evolution code \citep[\texttt{GENEC,}][]{eggenberger2008} to compute a grid of non-rotating MS models with solar metallicity ($Z=0.014$).
The masses included are 1.5, 2.5, 8, 15, 25, 32, 40 and 60\,M$_{\odot}$.
For each mass we compute at least one standard CBM model and one entrainment model. 

The standard CBM prescription in \texttt{GENEC} is step overshoot, where the convective core is extended by some distance $\alpha_{\rm{ov}}H_{P,\mathrm{b}}$.
In \texttt{GENEC}, the default value for $\alpha_{\mathrm{ov}}$ is 0.1 for models with initial mass $M_{\mathrm{ini}}\geq1.7$\,M$_{\odot}$, 0.05 for $1.7$\,M$_{\odot}>M_{\mathrm{ini}} \geq 1.25$\,M$_{\odot}$ and 0 for $M_{\mathrm{ini}}<1.25$\,M$_{\odot}$.
These default values were calibrated using the MS width of low mass stars \citep[for details see][]{ekstrom2012}. 
As in the core, the CBM (a.k.a. overshoot) region is mixed instantaneously (for both chemical species and entropy) and uses the adiabatic temperature gradient.

Table \ref{tab:models} lists the models computed and their key properties. The first four columns of Table \ref{tab:models} define the initial parameters of the model. These are the initial mass $M_{\rm{ini}}$ and the CBM parameters (either $\alpha_{\rm{ov}}$ for step overshoot models or a combination of $A$ and $n$ for entrainment models).
The $\tau_{\rm{MS}}$ column is the main sequence lifetime. This is defined as the age of the model when the central hydrogen mass fraction has reached $10^{-4}$.
The next column, $T_{\rm{eff,min}}$, is the minimum effective temperature reached by the model during the MS. 
Next is the mean of the bulk Richardson number, $\langle Ri_{\rm B}\rangle$, taken over the duration of $\tau_{\rm{MS}}$, along with the means of its components, $\langle v_{\rm c} \rangle$ and $\langle l_{\rm c} \Delta b \rangle$.
The final three columns pertain to the model attributes at the end of the MS.
These include the final mass, $M_{\rm{fin}}$, the mass of the helium core, $M_{\rm{He}}$, and the total mass entrained, $M_{\rm{ent,tot}}$.
$M_{\rm{He}}$ is defined as the mass of the convective core at a central hydrogen mass fraction of one per cent.

Both a default step overshoot model and an entrainment model with $A=10^{-4}$ and $n=1$ were calculated for each mass.
This value of $A$ was chosen to reproduce the MS lifetime of the 2.5\,M$_{\odot}$ standard overshoot model, as this mass is within the mass range originally used to calibrate the step overshoot. 
$A=2\e{-4}$ was also used for some masses to explore the widening of the MS in the high-mass range.

Previous simulations of convection have found that $n\sim1$, which guided our choice to keep $n=1$ for the majority of our grid. However, the $A$ values used for our 1D MS models ($A\sim10^{-4}$) are substantially lower than those derived from 3D simulations. 
$A$ values derived from 3D simulations include $A=1.06$ \citep[][oxygen burning]{meakin2007}, $A\approx0.1$ \citep[][oxygen burning]{muller2016} and $A=0.05$ \citep[][carbon burning]{cristini2019}. The difference could simply be a matter of evolutionary phase, since these 3D simulations are all of later stages than the MS. One potential confounding factor is radiative diffusion. Since the burning stages from carbon onward are neutrino-cooled, the effect of radiative diffusion on the mixing process is minimal, in contrast to the MS. Another point is partial degeneracy, which plays a part in later-stage stellar evolution but not in MS convective cores. Finally, the entrainment law may not keep the same slope for all $Ri_{\rm B}$ values. Our 1D models have $Ri_{\rm B}$ in the range of $\sim$10$^4$ to $\sim$10$^7$, which is substantially higher than the upper limit of $Ri_{\rm B}\sim1000$ in the 3D simulations and may represent a different entrainment law regime. Alternatively, there may be other important physics which is not encompassed by the entrainment law in its current form.

See Section\,\ref{sec:parameters} for more details on the chosen entrainment parameter values.
Appendix \ref{app} contains details on model resolution.

\begin{table*}
 \centering
 \caption{Summary of the CBM parameters used in the grid along with some key quantities. See Section\,\ref{sec:grid} for a description of the columns.}
 \label{tab:models}
 \begin{tabular}{cccccccccccc}
  \hline
  $M_{\rm{ini}}$ & $\alpha_{\rm{ov}}$ & $A$ & $n$ & $\tau_{\rm{MS}}$ & $\lg{(T_{\rm{eff,min}})}$ & $\lg{\langle Ri_{\rm B}\rangle}$ & $\lg{\langle v_{\rm c}\rangle}$ & $\lg{\langle l_{\rm c}\Delta b\rangle}$ &$M_{\rm{fin}}$ & $M_{\rm{He}}$ & $M_{\rm{ent,tot}}$\\
  {[M$_{\odot}$]} & & & & {[Myr]} & {[K]} & & {[cm\,s$^{-1}$]} & {[cm$^2$\,s$^{-2}$]} & {[M$_{\odot}$]} & {[M$_{\odot}$]} & {[M$_{\odot}$]}\\
  \hline
  1.5 & 0.05 & - & - & 2093 & 3.82 & 7.54 & 3.29 & 14.2 & 1.50 & 0.0662 & -\\
  1.5 & - & $10^{-4}$ & 1 & 2060 & 3.82 & 7.60 & 3.29 & 14.2 & 1.50 & 0.0723 & 0.0135\\
  \\
  2.5 & 0.1 & - & - & 512 & 3.93 & 6.84 & 3.66 & 14.2 & 2.50 & 0.173 & -\\
  2.5 & - & $5\e{-5}$ & 1 & 486 & 3.94 & 6.87 & 3.66 & 14.2 & 2.50 & 0.181 & 0.0395\\
  2.5 & - & $10^{-4}$ & 1 & 519 & 3.92 & 6.86 & 3.66 & 14.2 & 2.50 & 0.227 & 0.0869\\
  2.5 & - & $2\e{-4}$ & 1 & 582 & 3.90 & 6.86 & 3.67 & 14.2 & 2.50 & 0.319 & 0.199\\
  2.5 & - & $3\e{-4}$ & 1 & 668 & 3.87 & 6.89 & 3.68 & 14.3 & 2.50 & 0.457 & 0.368\\
  \\
  8 & 0.1 & - & - & 31.8 & 4.27 & 5.76 & 4.22 & 14.2 & 8.00 & 0.933 & -\\
  8 & - & $10^{-4}$ & 1 & 33.1 & 4.26 & 5.80 & 4.23 & 14.3 & 8.00 & 1.25 & 0.403\\
  8 & - & $2\e{-4}$ & 1 & 36.5 & 4.24 & 5.82 & 4.23 & 14.3 & 8.00 & 1.69 & 0.833\\
  \\
  15 & 0.1 & - & - & 11.6 & 4.39 & 5.29 & 4.46 & 14.2 & 14.8 & 2.82 & -\\
  15 & 0.3 & - & - & 13.0 & 4.35 & 5.27 & 4.47 & 14.2 & 14.7 & 3.69 & -\\
  15 & 0.5 & - & - & 14.3 & 4.31 & 5.24 & 4.47 & 14.2 & 14.7 & 4.55 & -\\
  15 & - & $10^{-4}$ & 1 & 12.3 & 4.37 & 5.34 & 4.47 & 14.3 & 14.8 & 3.72 & 0.960\\
  15 & - & $2\e{-4}$ & 1 & 13.8 & 4.34 & 5.38 & 4.47 & 14.3 & 14.7 & 5.01 & 2.06\\
  15 & - & $10^{-4}$ & 0.9 & 15.1 & 4.27 & 5.49 & 4.48 & 14.5 & 14.6 & 6.24 & 3.24\\
  15 & - & $10^{-4}$ & 1.2 & 10.9 & 4.40 & 5.40 & 4.46 & 14.3 & 14.8 & 2.52 & 0.0881\\
  15 & - & $10^{-4}$ & 1.5 & 10.9 & 4.40 & 5.28 & 4.46 & 14.2 & 14.8 & 2.39 & 0.00350\\
  \\
  25 & 0.1 & - & - & 6.54 & 4.43 & 5.00 & 4.62 & 14.2 & 24.2 & 6.64 & -\\
  25 & 0.3 & - & - & 7.14 & 4.37 & 4.97 & 4.62 & 14.2 & 23.8 & 8.14 & -\\
  25 & 0.5 & - & - & 7.70 & 4.25 & 4.95 & 4.63 & 14.2 & 23.0 & 9.54 & -\\
  25 & 0.7 & - & - & 8.18 & 3.96 & 4.94 & 4.63 & 14.2 & 20.4 & 10.8 & -\\
  25 & - & $10^{-4}$ & 1 & 6.99 & 4.39 & 5.08 & 4.62 & 14.3 & 24.1 & 8.73 & 1.90\\
  25 & - & $2\e{-4}$ & 1 & 7.63 & 4.31 & 5.11 & 4.63 & 14.4 & 23.3 & 10.9 & 3.72\\
  \\
  32 & 0.1 & - & - & 5.30 & 4.43 & 4.85 & 4.68 & 14.2 & 30.1 & 9.55 & -\\
  32 & 0.3 & - & - & 5.72 & 4.32 & 4.82 & 4.69 & 14.2 & 28.9 & 11.4 & -\\
  32 & 0.5 & - & - & 6.09 & 3.78 & 4.80 & 4.69 & 14.2 & 24.9 & 13.1 & -\\
  32 & - & $10^{-4}$ & 1 & 5.66 & 4.33 & 4.92 & 4.69 & 14.3 & 29.1 & 12.4 & 2.49\\
  32 & - & $2\e{-4}$ & 1 & 6.08 & 4.00 & 4.94 & 4.69 & 14.3 & 25.3 & 15.4 & 4.87\\
  \\
  40 & 0.1 & - & - & 4.51 & 4.40 & 4.71 & 4.73 & 14.2 & 36.5 & 13.0 & -\\
  40 & 0.3 & - & - & 4.84 & 3.88 & 4.69 & 4.74 & 14.2 & 30.0 & 15.2 & -\\
  40 & 0.5 & - & - & 5.11 & 3.83 & 4.67 & 4.74 & 14.1 & 24.6 & 17.1 & -\\
  40 & - & $10^{-4}$ & 1 & 4.86 & 3.63 & 4.78 & 4.74 & 14.3 & 29.5 & 17.1 & 3.35\\
  \\
  60 & 0.1 & - & - & 3.58 & 4.08 & 4.63 & 4.74 & 14.1 & 36.6 & 21.2 & -\\
  60 & 0.3 & - & - & 3.79 & 4.23 & 4.59 & 4.75 & 14.1 & 36.4 & 24.8 & -\\
  60 & 0.5 & - & - & 3.95 & 4.28 & 4.57 & 4.75 & 14.1 & 37.8 & 28.0 & -\\
  60 & - & $10^{-4}$ & 1 & 3.75 & 4.10 & 4.66 & 4.75 & 14.2 & 34.6 & 26.2 & 3.51\\
  \hline
 \end{tabular}
\end{table*}

\section{Results}
\label{sec:results}
\subsection{Time dependence of boundary penetrability and mass entrainment rate}
\label{sec:time_dependence}
\begin{figure}
 \includegraphics[width=\columnwidth]{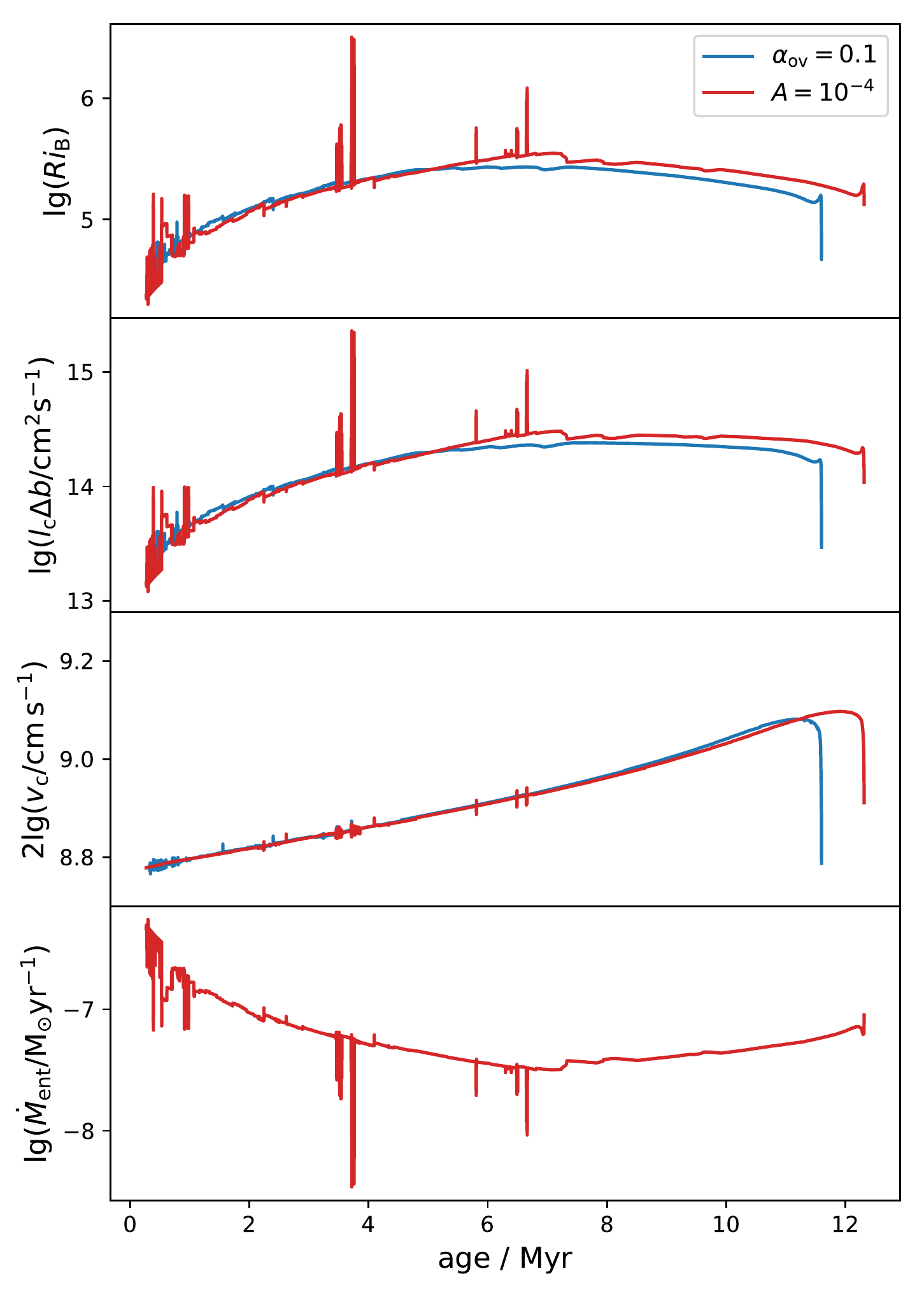}
 \caption{Time evolution of the bulk Richardson number for 15\,M$_{\odot}$ models with either the default step overshoot parameter $\alpha_{\rm{ov}}=0.1$ or entrainment parameters $A=10^{-4}$ and $n=1$. From top to bottom, the figure presents the bulk Richardson number, $Ri_{\rm B}$, the buoyancy jump multiplied by the length-scale for turbulent motions, $l_{\rm c}\Delta b$ (the numerator of $Ri_{\rm B}$), the mass-weighted mean square of the MLT velocity, $v_{\rm c}$, throughout the convective region (the denominator of $Ri_{\rm B}$) and finally the corresponding mass entrainment rate, $\dot{M}_{\rm{ent}}$ (using Eq.\,\ref{eq:mdot}).}
 \label{fig:15dov0p1rib}
\end{figure}

The time dependence of the bulk Richardson number, $Ri_{\rm B}$, for two 15\,M$_{\odot}$ models (step overshoot with $\alpha_{\rm{ov}}=0.1$, entrainment with $A=10^{-4}$ and $n=1$) is presented in Fig.\,\ref{fig:15dov0p1rib}. Over the MS, the variations in $Ri_{\rm B}$ are modest, within one order of magnitude. Nevertheless, we can see in Fig.\,\ref{fig:15dov0p1rib} that $Ri_{\rm B}$ initially increases and later on decreases. 

\begin{figure*}
 \includegraphics[width=\textwidth]{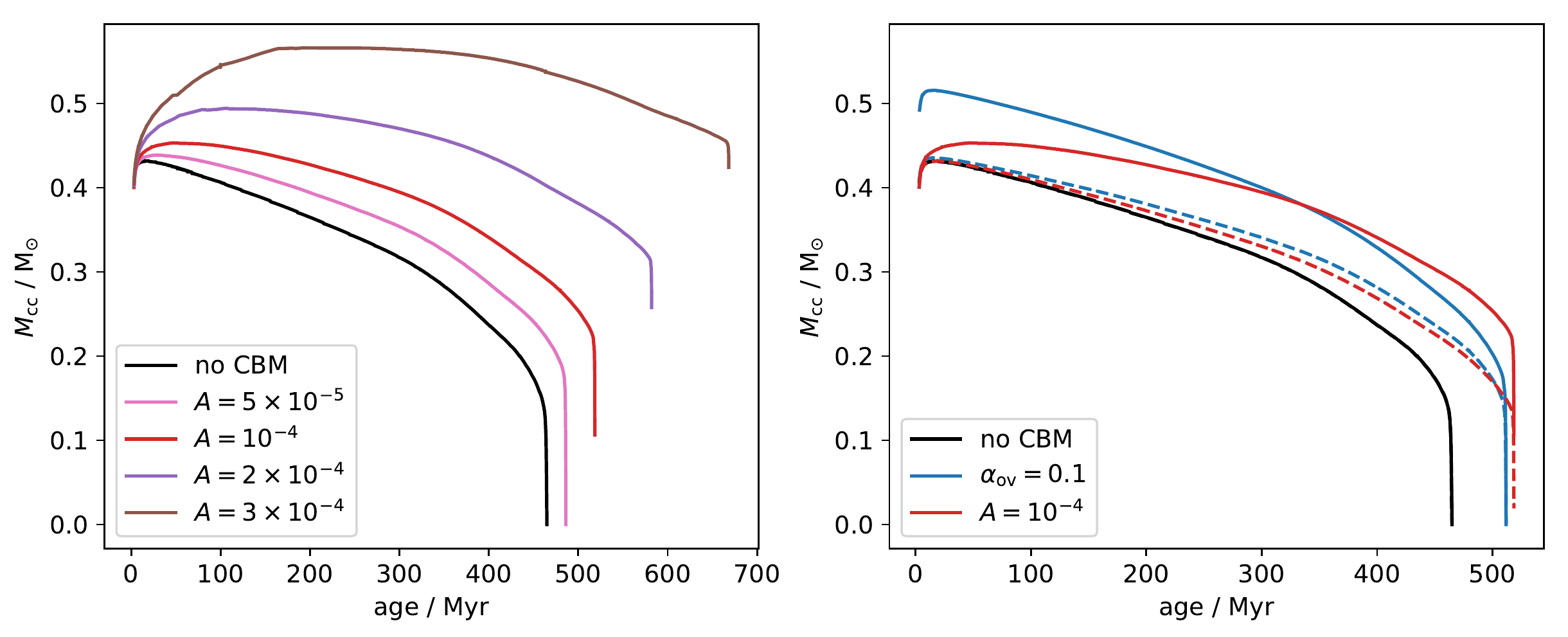}
 \caption{Comparison of the evolution of the convective core mass (Schwarzschild convective region plus CBM region), $M_{\rm cc}$, for 2.5\,M$_{\odot}$ models with step overshoot $\alpha_{\rm ov}=0.1$ and entrainment. Entrainment models use $n=1$. \textit{Left:} Time evolution of $M_{\rm cc}$  for various values of $A$, including $A=0$ (no CBM). \textit{Right:} The step overshoot model compared to the entrainment model with the closest-matching lifetime ($A=10^{-4}$) and the model without CBM. The dashed lines show the mass coordinate of the Schwarzschild boundary. The entire convective region and CBM region (for both overshooting and entrainment) are assumed to be fully mixed (for both chemical species and entropy) and have an adiabatic temperature gradient.}
 \label{fig:acompare}
\end{figure*}

The increase in $Ri_{\rm B}$ can be understood by considering the evolution of the buoyancy jump $\Delta b$ (the length-scale for turbulent motions, $l_{\rm c}$, which is set to half of a pressure scale height, is roughly constant during the MS), which is an integration of the buoyancy frequency $N^2$ over the boundary region.
In a massive star such as the 15\,M$_{\odot}$ model plotted, the convective core continuously recedes in mass over the MS. As the convective core recedes, it leaves behind a chemical gradient which contributes to an increase in $N^2$ and hence $\Delta b$. This leads to an increase in $l_{\rm c}\Delta b$ (the numerator in $Ri_{\rm B}$ shown in the second row of Fig.\,\ref{fig:15dov0p1rib}), which is strongest at the very beginning of the main sequence since there is no chemical composition gradient to start with.
After some time  (age $\sim 6.5\,$Myr), the core recedes far enough that the outermost limit 
of the buoyancy jump integration 
is lower than the original extent of the convective core.
From this point onward, $\Delta b$ remains roughly constant
since the full extent of its integration region is already occupied by the chemical gradient left by the convective core. Note that this saturation would likely occur earlier in the evolution if the size of the integration region was smaller; see the text below Eq.\,\ref{eq:buoyjump} in Section\,\ref{sec:calc_rib}.
The transient spikes in $Ri_{\mathrm{B}}$ also come from spikes in $\Delta b$. These originate from the finite differencing used in the code, since the boundary lies between two grid points. Fortunately, they have no impact on the results since they cause a temporary decrease in the entrainment mass rate (bottom row in Fig.\,\ref{fig:15dov0p1rib}). The spikes can be further explained by considering the integration of the squared buoyancy frequency. The buoyancy frequency depends on the gradient of the mean molecular weight, $\nabla_{\rm \mu}$ (see Eq.\,\ref{eq:buoyfreq}), which becomes the dominant part of $N^2$ at the upper edge of the CBM region. In the absence of mixing above this edge, $\nabla_{\rm \mu}$ can experience large local spikes. This is reflected in the mean molecular weight $\mu$ as step-like features rather than a smooth profile, and can cause transient increases in $Ri_{\rm B}$. These perturbations in $Ri_{\rm B}$ do not cause pathological changes in the core mass, which evolves smoothly (see Fig.\,\ref{fig:acompare}).

We intentionally did not include any shear mixing beyond the entrained region to study the effects of entrainment without any additional extension the MS lifetime from other factors. Any additional mixing processes such as shear would make it difficult to determine how much the entrainment itself affects the MS width and lifetime. However, we know from 3D simulations that there is a shear layer, which will smooth composition and structure profiles and probably prevent these spikes in the models \citep{arnettmoravveji2017,jones2017}. This shear layer could be modelled using an exponentially decaying diffusion coefficient \citep[exp-D hereinafter,][]{freytag1996,herwig2000} at the edge of the entrained region. Preliminary results suggest that a combination of entrainment and exp-D improves the transient spikes in $Ri_{\rm B}$ seen in pure entrainment models. Since exp-D provides an extra source of CBM, smaller values of $A$ might be needed in these combination models to produce the required MS widths.

\begin{figure*}
        \includegraphics[width=\textwidth]{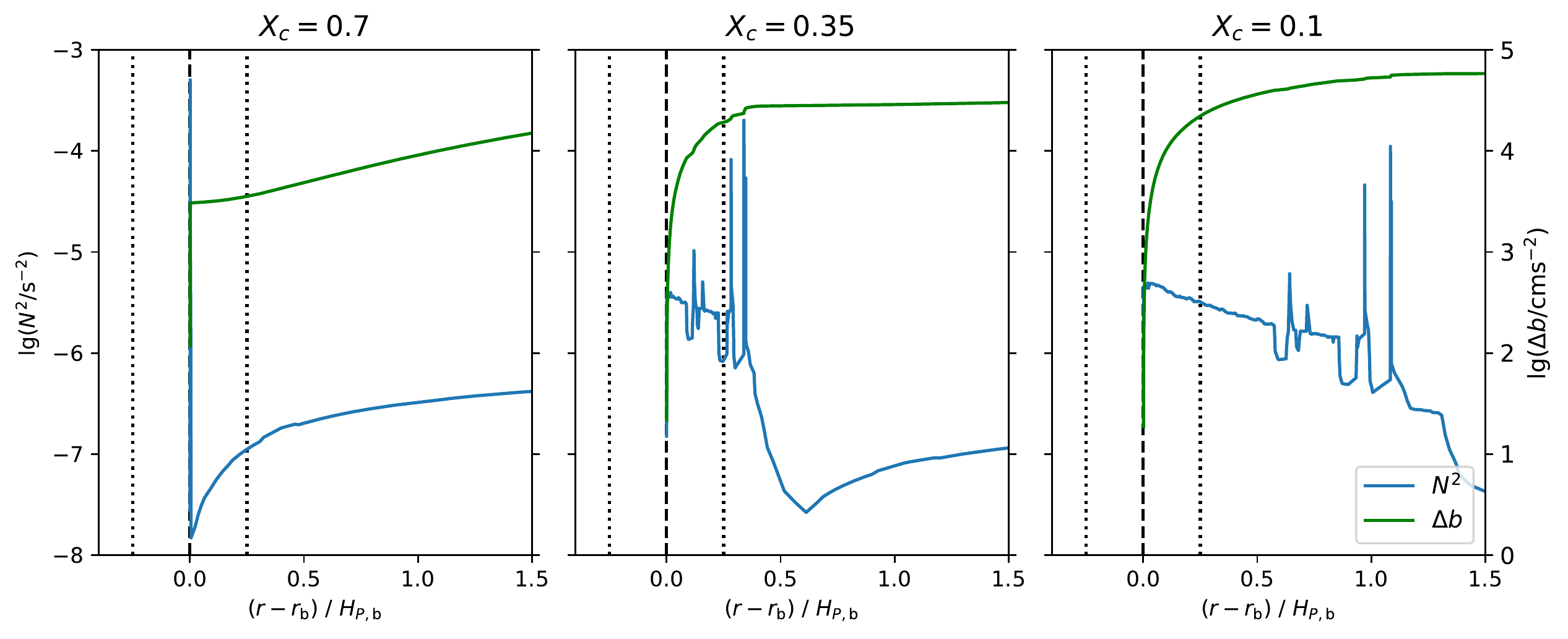}
    \caption{Profiles of the buoyancy frequency $N^2$ \textit{(blue)} at three different central hydrogen mass fractions, $X_{\rm c}$ (indicated at the top of the panel), in a 15\,M$_{\odot}$ entrainment model with $A=10^{-4}$ and $n=1$. Also shown is the buoyancy jump $\Delta b$ \textit{(green, right axis)} when integrated out to the corresponding number of pressure scale heights from the boundary, shown on the x-axis. The dashed line at $(r-r_{\rm b})/H_{P,\rm{b}}=0$ is the border between the entrained region and the stable region at $r=r_{\rm b}$. The dotted lines represent the limits of the buoyancy jump integration as used in our models ($r_{\rm b} \pm 0.25 H_{P,\rm b}$).}
    \label{fig:N2profiles}
\end{figure*}

Figure\,\ref{fig:N2profiles} shows the buoyancy jump integration region at three stages of the evolution of the 15\,M$_{\odot}$ entrainment model with $A=2\e{-4}$. The dashed lines represent the position of the edge of the entrained region at $r=r_{\rm b}$. The dotted lines represent the upper and lower limits of the integration. Both $N^2$ \textit{(blue)} and $\Delta b$ \textit{(green)} are plotted, with $\Delta b$ being the value obtained when integrating from the convective boundary to the corresponding radius on the x-axis. Hence, the value for $\Delta b$ at $(r-r_{\rm b})/H_{P,\rm b}=0.25$ is the value plotted in Fig.\,\ref{fig:15dov0p1rib}. Values for $\Delta b$ at higher radius would be obtained if the upper limit of the integration was larger.

Since the temperature gradient in the entrained region is adiabatic, $N^2$ is only positive in the stable region, which is the only region to contribute to the buoyancy jump in our current models. Fig.\,\ref{fig:N2profiles} also shows that the main contribution to the buoyancy jump is from the region close to $r=r_{\rm b}$. Outside our chosen integration region, $\Delta b$ remains at a similar order of magnitude. If the integration region is in fact significantly smaller than our chosen value, e.g. $0.05 H_P$, then the buoyancy jump would also be significantly smaller.

Finally, the modest decrease in $Ri_{\mathrm{B}}$ towards the end of the MS is due to the gradual increase in convective velocities (third row in Fig.\,\ref{fig:15dov0p1rib}). The increase in convective velocities is due to the luminosity of the star gradually increasing over the MS. Since velocity and luminosity are related by $v_c^3 \propto L$ \citep{biermann1932}, convective velocities also increase over the MS. Compared to $\Delta b$, however, the variation in $v_{\mathrm{c}}$ is small, which explains why $\Delta b$ has the greatest effect on the overall changes of $Ri_{\mathrm{B}}$ during the MS.

Over the MS, $Ri_{\rm B}$ varies between a few tens of thousands and a few hundreds of thousands (excluding short spikes explained above). Using the entrainment law (Eq.\,\ref{eq:entrainmentlaw}) with $A=10^{-4}$ and $n=1$, this leads to mass entrainment rates between $10^{-6.3}$ and $10^{-7.5}$\,M$_\odot\,$yr$^{-1}$ in a 15\,M$_{\odot}$ model. 
The mass entrainment rate, which is inversely proportional to $Ri_{\rm B}$, first decreases during the first part of the MS and later on increases slightly. The mass entrainment rate in this model leads to a total entrained mass of 0.960\,M$_\odot$ (see last column of Table\,\ref{tab:models}).



\subsection{Mass dependence of boundary penetrability}
\label{sec:mass_dependence}
Current observations seem to suggest that convective boundary mixing is mass dependent.
For instance, \citet{claret2019} presented the dependence of CBM as a function of mass for stars of less than $\sim 4$\,M$_{\odot}$ in binary systems, finding a steep dependence for the lowest mass stars with growing convective cores on the MS. \citet{schootemeijer2019} found a mild dependence of CBM on mass for stars in the Small Magellanic Cloud. \citet{higgins2019} compared models to the massive star binary HD 166734, concluding that a step overshoot parameter of $\alpha_{\rm{ov}}=0.5$ was suitable for stars above 30 to 40\,M$_\odot$, which is much larger than the value of $\alpha_{\rm{ov}}=0.1$ determined for lower mass stars by \citet{ekstrom2012}. \citet{castro2014} performed a large study on Milky Way stars and found significant broadening of the MS at higher masses; we compare to this work in particular in Section\,\ref{sec:MS_width}.
In this section, we explore if this dependence can be explained by the mass dependence of stellar structure and properties.

It is well known that the luminosity has a strong mass dependence.
For low-mass stars, the dependence is steep with $L \propto M^3$. For massive stars, it flattens and approaches a linear dependence with mass above about 20\,M$_\odot$ (see Fig.\,6 in \citealt{yusof2013}). A higher luminosity leads to higher convective velocities ($v_c^3 \propto L$, \citealt{biermann1932}). Since the bulk Richardson number, $Ri_{\rm B}$, contains a velocity term, it would also be expected to show mass dependence.

\begin{figure*}
        \includegraphics[width=\textwidth]{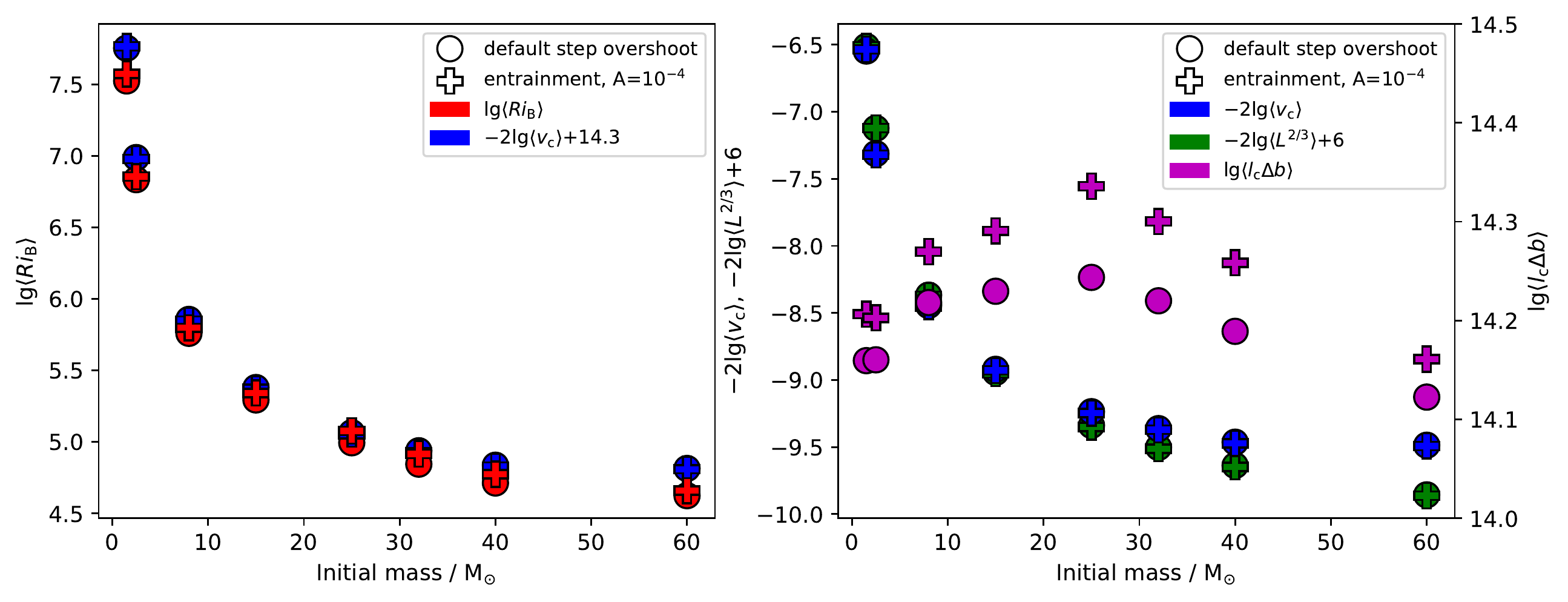}
    \caption{
Mass dependence of the Bulk Richardson number and its components. 
    The left hand panel shows the time-average of the log$_{10}$ of the bulk Richardson number over the MS against initial mass. 
    Circles represent default step overshoot models ($\alpha_{\rm{ov}}=0.05$ for 1.5\,M$_{\odot}$, $\alpha_{\rm{ov}}=0.1$ otherwise) and pluses represent entrainment models with $A=10^{-4}$ and $n=1$.
    The right panel shows the log$_{10}$ of the time average of the two components of $Ri_{\rm B}$. For the denominator ($v_c^2$), a minus sign is used so that adding the values of the two components yields the value of $Ri_{\rm B}$.
    }
    \label{fig:RiB_components}
\end{figure*}

The left panel of Figure \ref{fig:RiB_components} shows the logarithm of the time average of two values: $Ri_{\rm B}$ and $v_{\rm c}^2$ (sign reversed, since it is the denominator of $Ri_{\rm}$, and scaled by a constant value to fit on the same axis). This panel demonstrates that $Ri_{\rm B}$ is mass dependent and its dependence is dominated by the velocity term. The right panel shows the velocity term compared to total luminosity (again scaled by a constant), demonstrating that the mass dependence of velocity is also very similar to that of luminosity, as expected from the mass luminosity relation. Conversely, the buoyancy jump term also plotted in the right-hand panel does not demonstrate mass dependence since its logarithm varies by less than 0.5 dex. Despite this, the buoyancy jump term does dominate the variation of mass entrainment rate with time (Fig.\,\ref{fig:15dov0p1rib}) and so cannot be ignored when considering entrainment at the convective boundary. Note also that this only holds if our assumptions on the buoyancy jump integration region (see text below Eq.\,\ref{eq:buoyjump} in Section\,\ref{sec:calc_rib}) are correct.

In this section, we showed that convective boundary properties have a clear mass dependence, which can be measured via $Ri_{\mathrm{B}}$. Next, we want to explore whether the entrainment law, which uses $Ri_{\mathrm{B}}$ can provide the mass dependence of the convective boundary mixing needed to reproduce the observed MS width.
We can already note that $Ri_{\mathrm{B}}$ decreasing with initial mass will lead to higher entrainment rates for more massive stars, which goes in the right direction.


\subsection{Dependence of entrainment on the entrainment law parameters}
\label{sec:parameters}
Both 3D simulations and theoretical studies determined various values for the entrainment law parameters $A$ and $n$.
From a theoretical energy balance argument, $n$ should be 1 \citep{stevens2001}.
Hydrodynamical simulation values for $n$ range from $\sim0.7$ to $\sim1$ depending on the setup.
Conversely, literature values for $A$ vary from $A\approx1$ \citep{meakin2007} to $A\approx0.05$ \citep{cristini2019}.
See \citet{muller2020} and references therein for examples of entrainment law parameters derived from 3D simulation results.
The fact that $A$ and $n$ are not the same between setups suggests that the entrainment law in its current form does not encompass every aspect of the growth of the convective region in these simulations.

In this study, we start by taking $n=1$ and use published 1D \texttt{GENEC} evolution models with step overshoot and $\alpha_{\rm ov}=0.1$ to determine a value of $A$ that would reproduce the published models.
The value of $\alpha_{\rm ov}=0.1$ in \texttt{GENEC} models is constrained using the main sequence width for low/intermediate-mass stars \citep{ekstrom2012}.
The same value of $\alpha_{\rm ov}$ is then applied to all higher masses (at all metallicities) in the published grids of \texttt{GENEC} models.
Therefore, 2.5\,M$_{\odot}$ models were used to constrain an $A$ value in entrainment models that matches the general properties of the 2.5\,M$_{\odot}$ \texttt{GENEC} model with step overshoot and $\alpha_{\rm ov}=0.1$: MS width in the Hertzsprung-Russell (HR) diagram, core masses and MS lifetime.
Table\,\ref{tab:models} and Fig.\,\ref{fig:acompare} show the comparison between entrainment models with different values of $A$ and the default overshoot model. They confirm that the minimum effective temperatures reached by the models with $\alpha_{\rm ov}=0.1$ and $A=10^{-4}$, $n=1$ are very similar. Table\,\ref{tab:models} also indicates that the MS lifetimes are similar.


Figure \ref{fig:acompare} shows the evolution of the convective core mass in 2.5 M$_{\odot}$ entrainment models. The left-hand panel shows how the entrainment depends on the value of $A$ with values of $A$ ranging from zero (no CBM) to $3\times 10^{-4}$ (all models with $n=1$). As expected, a larger value of $A$ leads to more entrainment and thus larger convective core masses and longer lifetimes. One point to note is that since entrainment rate is reduced if $Ri_{\rm B}$ increases, the potential problem of the convective region quickly encompassing the whole star can be avoided. Indeed, as the entrainment extends further, the jump in composition and entropy at the boundary increases and makes the boundary stiffer, which makes it harder for additional entrainment. The use of the entrainment law thus provides an important feedback. This is best seen for the $A=3\times 10^{-4}$ model (brown curve), where entrainment leads to core growth only during the first part of the MS. After a while, the entrained mass plateaus since the entrainment rate drops significantly and the convective regions shrinks in mass due to the Schwarzschild boundary receding as in the step overshoot models.
Note that much larger values of $A$ may still lead to the entire model becoming convective. Much larger values of $A$ are not needed or supported by observations anyway as discussed in Section\,\ref{sec:MS_width}.

Keeping $n=1$, the value $A=10^{-4}$ provides the closest match to the default step overshoot model in terms of MS lifetime. We see, however, that the time evolution of the convective core is very different in entrainment models compared to the step overshoot model, as shown in the right-hand panel of Fig.\,\ref{fig:acompare}. The step overshoot model assumes that mixing is an instantaneous process (compared to the MS lifetime) and thus the convective core is significantly larger on the ZAMS in these models. On the other hand, 
entrainment is a time-dependent process (in that the size of the entrained region is dependent on the earlier entrainment history) and builds up over the entire MS, as shown in Eq.\,\ref{eq:ment}. The dashed-red line indicates the Schwarzschild boundary in the $A=10^{-4}$ model and shows how the entrained region (region between the solid and dashed red lines) grows in mass with time. This means that for a given lifetime, the entrainment models start with smaller and end with larger convective core masses compared to step overshoot models (see Table\,\ref{tab:models} and Section\,\ref{sec:core_masses}).

We also tested the dependence of entrainment on the value of $n$ with various 15\,M$_\odot$ models with values of $n=0.9, 1.2$ and 1.5 (keeping $A=10^{-4}$, see Table\,\ref{tab:models}). The dependence on $n$ is strong. Indeed, values of $n$ slightly larger than 1 (1.2 or 1.5) strongly reduce the total entrained mass (by a factor of 10 or more, see last column of Table\,\ref{tab:models}) and values of $n$ slightly smaller than 1 (0.9) lead to significantly more entrainment (by a factor of more than 3). While the dependence on $n$ and $A$ are not independent, our models tend to show that $n$ cannot be too far from 1.
We will compare the values determined in this study to observational constraints and hydrodynamic simulations in the discussion.


\begin{figure}
 \includegraphics[width=\columnwidth]{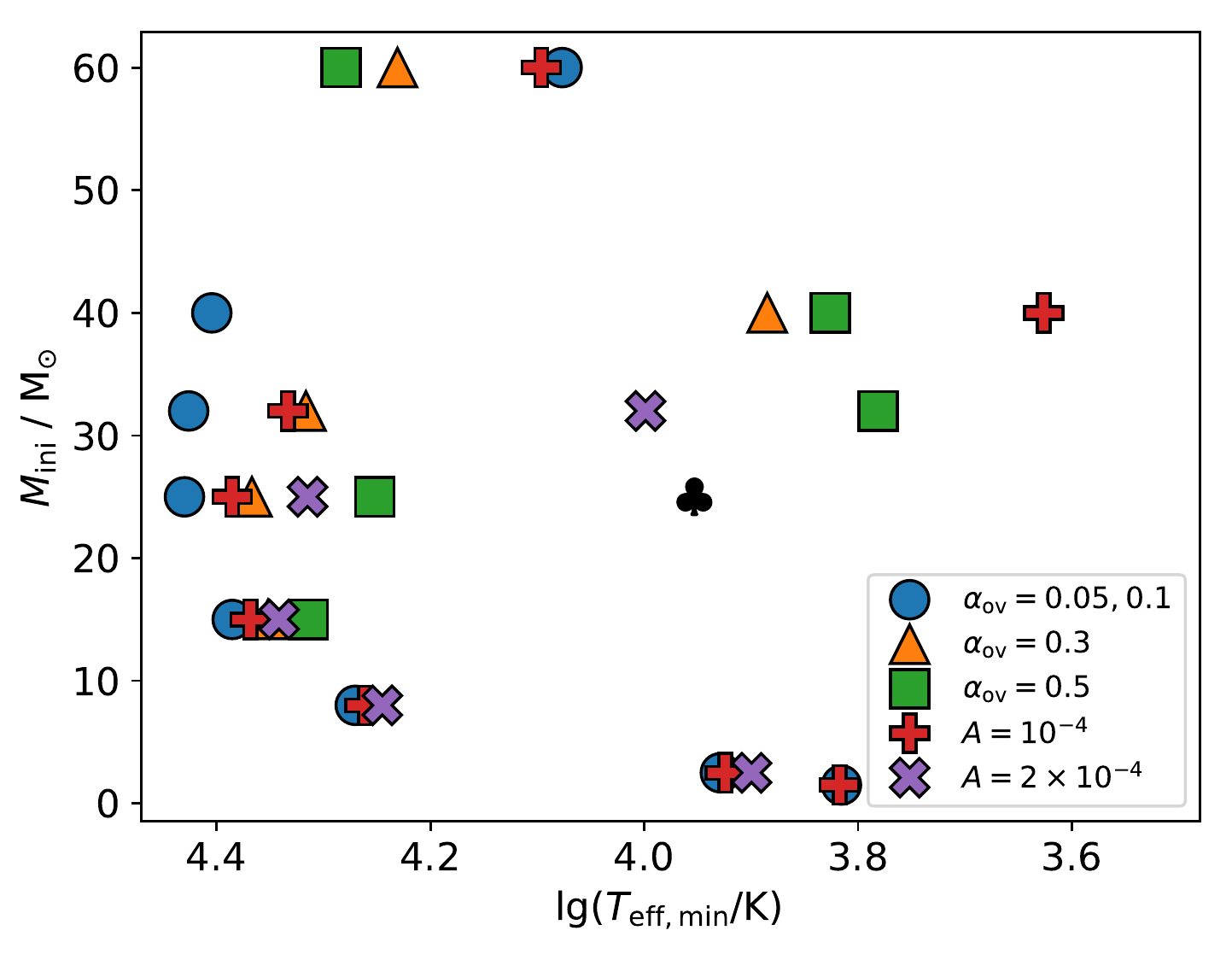}
 \caption{Minimum value of effective temperature on the main sequence for all models in the grid (as in Table\,\ref{tab:models}). The mixing schemes used are denoted by different coloured markers as in the legend. The one-off $\alpha_{\rm{ov}}=0.7$ model with 25\,M$_{\odot}$ is shown with the black clover symbol (see Section\,\ref{sec:discussion}).}
 \label{fig:minteff}
\end{figure}

\begin{figure*}
        \includegraphics[width=\textwidth]{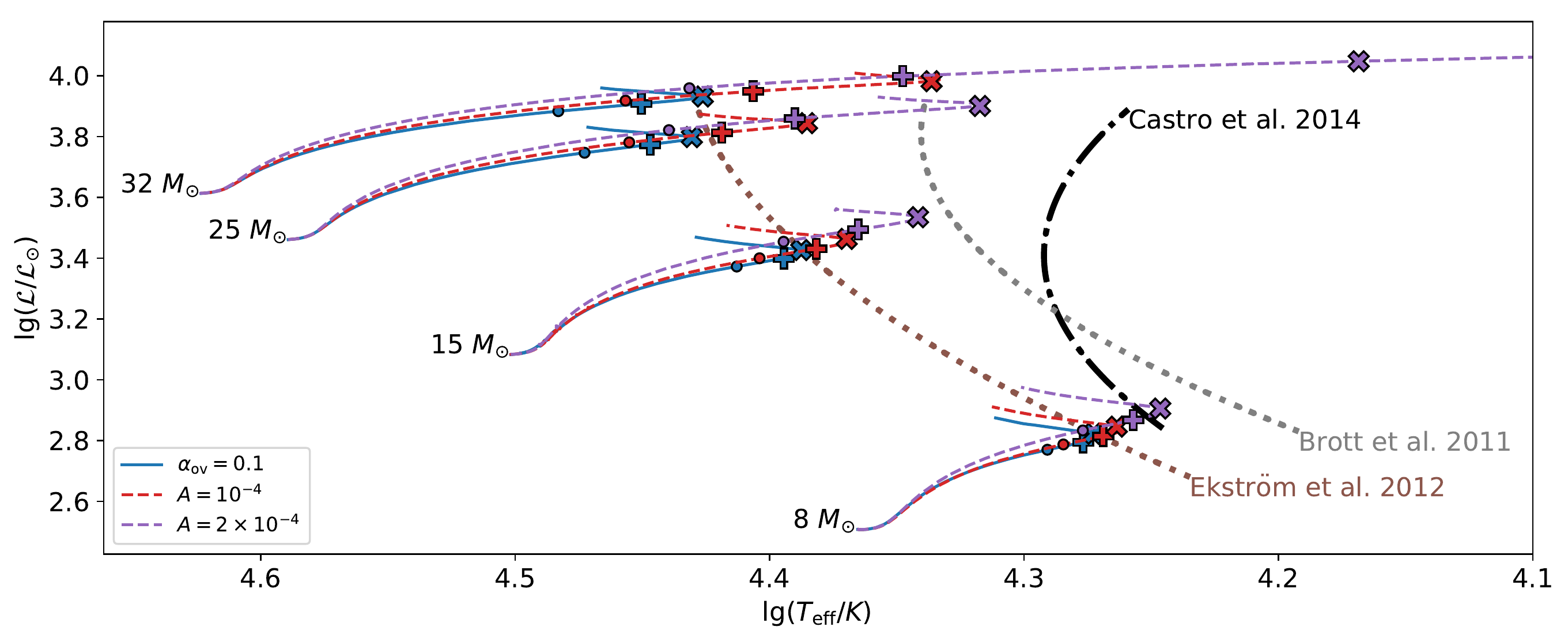}
    \caption{Spectroscopic HRD in the mass range 8 to 32\,M$_{\odot}$ with $\alpha_{\rm ov}=0.1$ step overshoot and two entrainment models, $A=10^{-4}$ and $2\e{-4}$. The dotted lines represent the position of the TAMS from model grids with $\alpha_{\rm ov}=0.1$ \citep{ekstrom2012} and $\alpha_{\rm ov}=0.335$ \citep{brott2011}. The dash-dotted line represents the position of the empirical TAMS determined by \citet{castro2014}; see their Table 1 for the polynomial coefficients of the three TAMS lines used in this figure. The dots, pluses and crosses have been placed where the model reaches 90\%, 95\% and 99\% of the MS lifetime respectively.}
    \label{fig:HRD_spec}
\end{figure*}

\subsection{Impact of entrainment on main sequence width}
\label{sec:MS_width}
One of the main observational constraints on stellar models is the MS width.
\citet{castro2014} represents one of the most comprehensive study of MS width at solar metallicity. One of their key findings is that models using a mass-independent value of step overshoot \citep{ekstrom2012,brott2011} do not reproduce the observed MS width. Instead, it appears that CBM must increase with initial mass. While the sample used in \citet{castro2014} is far from complete, it is worth comparing our new entrainment models with models with various amounts of step overshoot and the MS width deducted by \citet{castro2014}. \citet{castro2014} find that the MS generally extends to a $\lg{(T_{\rm eff})}\sim 4.3$ over a range of luminosities, which corresponds to stars in the mass range $\sim10-20$\,M$_{\odot}$. Above 20\,M$_{\odot}$, the MS does not seem to have a well-defined cool end and instead appears to extend down to very cool temperature.

Figure \ref{fig:minteff} shows the minimum effective temperature, $T_{\rm{eff,min}}$ reached on the MS by each model in the grid.
$T_{\rm{eff,min}}$ for the default step overshoot models is shown with blue disks and we can see that they indeed predict an MS cool edge that deviates from the observed $\lg{(T_{\rm{eff}})}\sim 4.3$ further as the mass of the model increases from 10\,M$_\odot$ upwards. We also see that these models do not predict the observed widening of the MS above 20\,M$_\odot$. As discussed in the previous section, entrainment models with $A=10^{-4}$ and $n=1$ (red pluses) reproduce the main features of the default ($\alpha_{\rm{ov}} = 0.1$) step overshoot models (MS lifetime and HRD tracks). Thus as expected, the $T_{\rm{eff,min}}$ of the entrainment models is also hotter than the observed one for stars between 10 and 20\,M$_{\odot}$. One difference appears for stars above 20\,M$_{\odot}$ with the entrainment models predicting a cooler edge for the 32\,M$_{\odot}$ model and a very cool edge for the 40\,M$_{\odot}$. The 60\,M$_{\odot}$ models do not follow this trend because they experience strong mass loss towards the end of the MS, which keeps the models on the hot side of the HRD. 

Increasing the value of $A$ from $10^{-4}$ to $2\times 10^{-4}$ (purple crosses) provides a reasonable match to the observed MS edge at 
$T_{\rm{eff,min}} \sim 4.3$. Indeed, $4.34 \leq T_{\rm{eff,min}} \leq 4.24$ in the models between 8 and 25\,M$_{\odot}$. Furthermore, the 32\,M$_{\odot}$ model now extends to very cool $T_{\rm{eff}}$.
While the observational constraints are not very tight, the MS width for lower masses is slightly wider than observations so a larger value of $A$ would not be favoured.
The broader MS width can be reproduced with an increased value of $\alpha _{\rm{ov}}$ (e.g. with $\alpha_{\rm{ov}} =0.5$, green squares) but in this case, the MS width for lower mass stars would be too wide.
The reason why the entrainment models have broader MS width for more massive stars is due to the mass dependence of $Ri_{\rm B}$ discussed in Section \ref{sec:mass_dependence}, which is used in the entrainment law. 
This means that the entrainment law provides a partial physical explanation for the apparent mass dependence of the overshooting parameters and a way of providing a much better fit to the observations with a single value of the parameters $A$ and $n$, which is harder for other CBM such as step overshoot or exponentially decaying diffusion coefficients. Whilst the fit to the TAMS edge could be improved, for example by varying $n$ in addition to $A$, the usefulness of this approach would be limited. Other factors such as rotation, metallicity variations and different mass loss prescriptions could provide additional mass-dependent factors which are not included in these models.

\citet{castro2014} gathered observations of galactic stars and placed them on a spectroscopic HRD (sHRD) \citep{langer2014}, in which they show the density of observed stars in each region of the HRD. Since the sample is incomplete and possibly biased \citep{vink2010,mcevoy2015}, it is difficult to compare densities of stars across the HRD. Nevertheless, it is still interesting to determine what fraction of the MS lifetime models spend in a given location in the HRD. This is indicated in Fig.\,\ref{fig:HRD_spec} (dots at  90\% of the MS lifetime, pluses at at 95\% and crosses at 99\%).


Figure \ref{fig:HRD_spec} shows our models on an sHRD so that we can compare directly to these observations. We focus on the 8 to 32\,M$_{\odot}$ range, which encompasses the region of the sHRD in which there is a clear observed TAMS boundary (marked on Fig.\,\ref{fig:HRD_spec} with the dash-dotted line). Also shown are the TAMS boundaries obtained by \citet{castro2014} from two model grids: \citet{ekstrom2012} using $\alpha_{\rm ov}=0.1$ and \citet{brott2011} using $\alpha_{\rm ov}=0.335$.

\begin{figure*}
 \includegraphics[width=\textwidth]{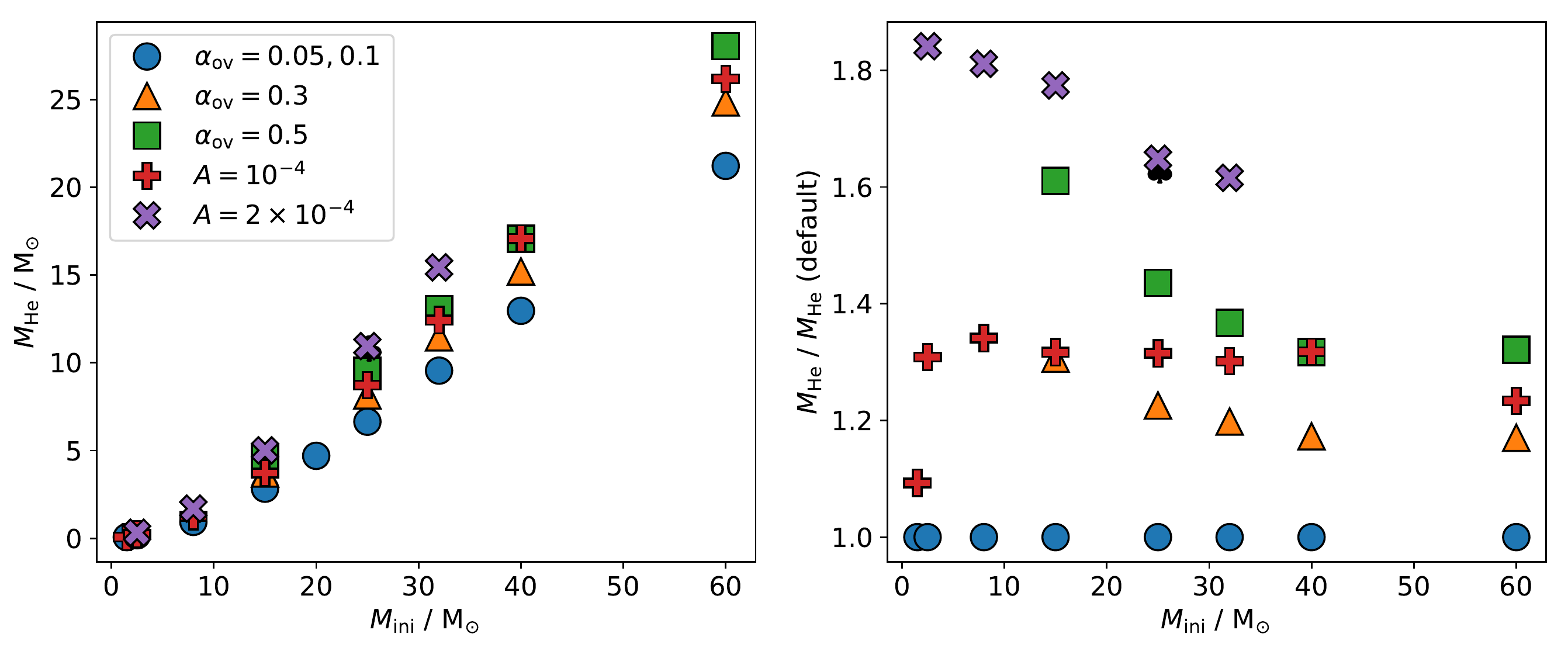}
 \caption{Final helium core mass, $M_{\rm He}$, for various values of step overshoot $\alpha_{\rm ov}$ and entrainment parameter $A$. All entrainment models use $n=1$. Blue circles represent the default value of $\alpha_{\rm ov}$, which is 0.05 for 1.5\,M$_{\odot}$ and 0.1 otherwise. The one-off $\alpha_{\rm{ov}}=0.7$ model with 25\,M$_{\odot}$ is shown with the black clover symbol, nearby the $A=2\e{-4}$ model (see Section\,\ref{sec:discussion}). \textit{Left:} Absolute value of $M_{\rm He}$ against initial mass. The $\alpha_{\rm ov}=0.1$ point at 20\,M$_{\odot}$ is taken from the \citet{ekstrom2012} grid. \textit{Right:} $M_{\rm He}$ normalised by the default step overshoot value.}
 \label{fig:MHe}
\end{figure*}

The \citet{ekstrom2012} step overshoot value of $\alpha_{\rm ov}=0.1$ was calibrated using models on the lower MS, such as our 2.5\,M$_\odot$ models. As such, its TAMS boundary is closest to the \citet{castro2014} empirical boundary in the lower mass range, but deviates strongly at higher masses. Conversely, the \citet{brott2011} step overshoot value of $\alpha_{\rm ov}=0.335$ was calibrated at 16\,M$_\odot$ and corresponds best to the \citet{castro2014} TAMS in the middle of the mass range, deviating at both the high-mass and low-mass extremes. This suggests that when using step overshoot to determine CBM, a mass-dependent $\alpha_{\rm ov}$ is 
needed in models to reproduce observations over a large mass range.

The entrainment law naturally accounts for the mass dependence of CBM through the mass dependence of $Ri_{\rm B}$ (see Section \ref{sec:mass_dependence}). In Fig.\,\ref{fig:HRD_spec}, the entrainment models have a markedly different TAMS boundary shape (approximated by the positions of the cross markers) with an increased widening of the MS with increasing mass. In particular, the $A=2\e{-4}$ model is closer to the \citet{castro2014} observed TAMS than both the \citet{ekstrom2012} and \citet{brott2011} TAMS boundaries for the 8 and 25\,M$_\odot$ models. Whilst the MS width for 32\,M$_\odot$ models is unconstrained by the \citet{castro2014} observations, the $A=2\e{-4}$ model does fulfill the requirement of reaching very low temperatures, with $\lg{T_{\rm eff}}\leq4.2$ at 99 per cent of the full MS lifetime.

\subsection{Impact of entrainment on helium core masses and lifetimes}
\label{sec:core_masses}
The type and degree of CBM 
also affects the mass of the helium core at the end of the MS. The size of this core, whilst not directly observable, has very important implications for post-MS evolution. The compactness and explodability of pre-supernova models is dependent on the post-MS structure, in which the helium core plays an important role \citep{oconnor2011,ertl2016,sukhbold2018,chieffi2020}.
Additionally, since the evolution is driven by the conditions in the core, CBM parameters that produce large cores can mimic the results of more massive models with less CBM.

In Table\,\ref{tab:models}, the helium core mass at the end of the MS, $M_{\rm He}$, is given in the penultimate column. We define $M_{\rm He}$ as the mass of the convective core (including CBM) when the central hydrogen mass fraction drops to one per cent. This definition gives similar results to taking the mass coordinate at which the hydrogen mass fraction drops to one per cent at the last time step of the MS.

Figure\,\ref{fig:MHe} shows both $M_{\rm He}$ (\textit{left}) and $M_{\rm He}$ divided by its value in the default step overshoot model (\textit{right}). As expected, the left panel shows that larger amounts of CBM produce larger core masses at the end of the MS. In absolute terms, this increase in core mass is greatest in the more massive stars. In the right-hand panel, the majority of CBM choices show the opposite trend, with a greater effect of CBM on relative core mass for the lower-mass models. This is particularly true for $\alpha_{\rm ov}=0.5$. In contrast, the $A=10^{-4}$ models  increase $M_{\rm He}$ by $\sim$30 per cent across the mass range of the grid, except for the 1.5\,M$_{\odot}$ model which displays a milder change in core mass.

The value of entrainment that best produces the \citet{castro2014} MS width in Fig.\,\ref{fig:HRD_spec} is $A=2\e{-4}$. As can be seen in the right-hand panel of Fig.\,\ref{fig:MHe}, this value of $A$ creates helium cores which are a factor of 1.6 to 1.8 larger than using default step overshoot models. The 20\,M$_{\odot}$ point \citep[taken from][]{ekstrom2012} in the left-hand panel of Fig.\,\ref{fig:MHe} illustrates the implications of this; a 15\,M$_{\odot}$ model with $A=2\e{-4}$ has a similar helium core mass to an $\alpha_{\rm ov}=0.1$ model, which is 5\,M$_{\odot}$ more massive initially. This shift of at least 5\,M$_{\odot}$ has wide-ranging implications for massive star evolution and their fate. Examples include the upper mass limit of observed supernova progenitors \citep{smartt2009} and the mass range of black hole production \citep[e.\,g.][]{chieffi2020}.

\begin{figure}
 \includegraphics[width=\columnwidth]{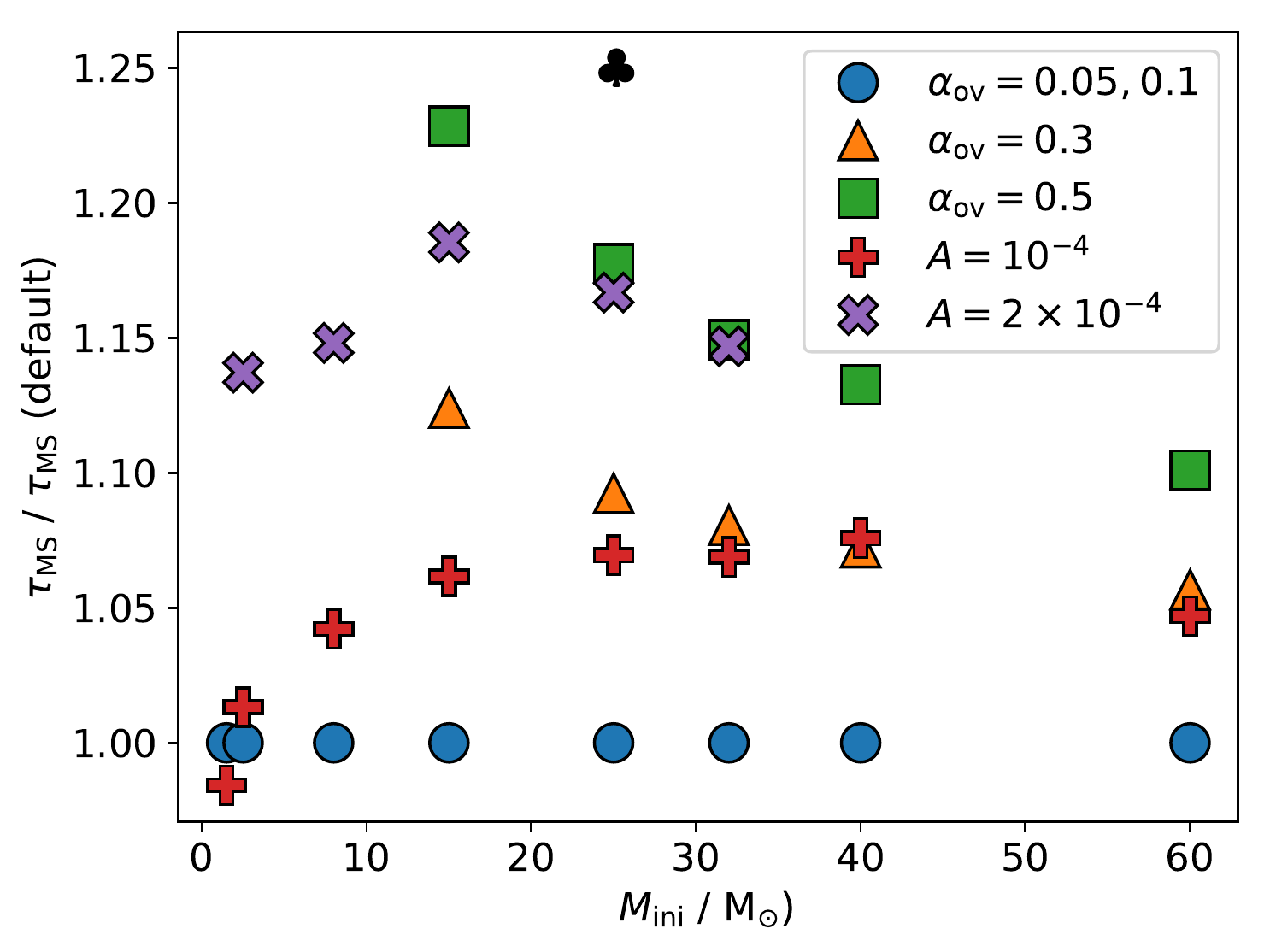}
 \caption{Main sequence lifetime relative to the default step overshoot models ($\alpha_{\rm ov}=0.05$ for 1.5\,M$_{\odot}$, 0.1 otherwise). The one-off $\alpha_{\rm{ov}}=0.7$ model with 25\,M$_{\odot}$ is shown with the black clover symbol (see Section\,\ref{sec:discussion}).}
 \label{fig:tauMS}
\end{figure}

Figure\,\ref{fig:tauMS} shows the MS lifetime, $\tau_{\rm MS}$ (column 5 in Table\,\ref{tab:models}), relative to the default step overshoot case for various CBM parameters across the mass range of the grid. This figure shows similar trends to the right-hand panel of Fig.\,\ref{fig:MHe}, but with more CBM producing longer lifetimes rather than larger cores. When comparing step overshoot models only, it is clear that the relative increase in lifetime is smaller for higher-mass stars. The entrainment models show more complicated non-monotonic behaviour. However, it is important to note that the relative effect of increasing CBM on lifetime is milder compared to the effect on helium core masses; the maximum relative increase in lifetime in Fig.\,\ref{fig:tauMS} is nearly 15 per cent for the 2.5\,M$_{\odot}$ model, whereas $M_{\rm He}$ is increased by a $\sim$80 per cent for the same model in Fig.\,\ref{fig:MHe}.

The strong effect on core masses and more modest effect on lifetimes can be understood from the difference between step overshoot and entrainment discussed in Section\,\ref{sec:parameters} and highlighted in Fig.\,\ref{fig:acompare}. Whilst the mass contained within the CBM region in the step overshoot models decreases with time, entrainment is a cumulative process, which builds up over the main sequence and thus leads to much larger final core masses.

Another important difference for the later evolution is the initial sizes of convective cores. The step overshoot model starts with a much larger core. This will leave an imprint on the structure of that part of the star, which will affect the behaviour of the intermediate convective zone \citep{kaiser2020}.

\section{Discussion and Conclusions}
\label{sec:discussion}


We have calculated a grid of 1D stellar models using the Geneva stellar evolution code with masses between 1.5 and 60\,M$_{\odot}$ and at solar metallicity ($Z=0.014$).
We have shown that the boundary penetrability by convective flows, quantified by the bulk Richardson number $Ri_{\rm B}$, decreases monotonically with increasing mass.
This decrease is dominated by the increase in typical convective velocities due to the steep mass-luminosity relation for stars in the 1 to 20\,M$_{\odot}$ range. 
The boundary stiffness, $l_{\rm c}\Delta b$, is nearly invariant with mass in the range studied.

Due to the decrease in $Ri_{\rm B}$ with mass, models with entrainment experience a mass-dependent increase in mixing.
This is reflected in a corresponding mass-dependent MS widening in the HRD.
For our models, we find the value of $A$ which best reproduces the observed MS widths of massive stars is $A=2\e{-4}$, with $n=1$. Note however that more extended samples are desired to place a very tight constraint on $A$ and that the effects of rotation were not considered in this work (Martinet et al. in prep.).


The choice of temperature gradient in the entrained region is also an important factor in the implementation of entrainment. As explained in Section\,\ref{sec:algorithm}, we use $\nabla=\nabla_{\rm ad}$ in the entrained region, since 3D simulations show that entropy is well mixed as the convective region grows \citep{cristini2017}. 3D simulations also show a narrow boundary above the entrained region with a smooth chemical gradient rather than a switch from one $\mu$ to another; it is likely that the mixing of entropy is similarly slowed compared to the entrained region in this boundary. Indeed, asteroseismic observations support MS convective cores with a smooth $\nabla$ profile in the CBM region \citep{arnettmoravveji2017}. 

In standard models, the global evolutionary effect of a slight change in $\nabla$ in the CBM region is subtle, especially if the CBM region is small. In entrainment models, however,  the size of the CBM region towards the end of the MS can be significant, especially with larger values of $A$. The choice of $\nabla$ may also have a more important role in entrainment models due to its effect on the buoyancy jump, $\Delta b$. In our current implementation, the CBM region has no contribution to $\Delta b$ whatsoever, since it is fully mixed ($\nabla_{\rm \mu}=0$) and $\nabla=\nabla_{\rm ad}$. This means that the buoyancy frequency in the entrained region is 0. If the temperature gradient were to instead transition smoothly from $\nabla_{\rm ad}$ to $\nabla_{\rm rad}$ within the entrained region \citep[as explored in][]{michielsen2019}, there would be a contribution to $\Delta b$ from the entrained region. This contribution would grow larger as the entrained region grows in size, therefore providing more feedback slowing the entrainment for larger values of $A$. Consequently, these models would require larger values of $A$ than models with $\nabla=\nabla_{\rm ad}$ to produce the same MS width.

Since we have demonstrated in Section\,\ref{sec:mass_dependence} that the mass dependence of $Ri_{\rm B}$ is dominated by the change in typical convective velocities with mass, it is interesting to test whether a scaling based on $v_{\rm c}$ could provide a simpler alternative to entrainment. This seems reasonable since
the dependence of $Ri_{\rm B}$ with mass is almost entirely controlled by $v_{\rm c}$, with $l_{\rm c}\Delta b$ staying nearly constant with mass. To constrain this scaling, we take the value of $\alpha_{\rm{ov}}=0.5$ for 15\,M$_{\odot}$, as this most closely matches the \citet{castro2014} observational $\lg{(T_{\rm{eff,min}})}\sim4.3$. The scaled overshoot parameter for each mass, $M_{\rm{ini}}$, is then given by
\begin{equation}
 \label{eq:scaledov}
 \alpha_{\rm ov,scaled,}(M_{\rm{ini}}) = \alpha_{15\,\rm M_{\odot}} \frac{\langle v^m(M_{\rm{ini}})\rangle}{\langle v^m(M_{\rm{ini}}=15\rm{\,M}_\odot)\rangle},
\end{equation}
where $\alpha_{15\,\rm M_{\odot}}=0.5$ and $\langle v^m(M_{\rm{ini}})\rangle$ is the average of the convective velocity to the power $m$ over the MS of the model of initial mass $M_{\rm{ini}}$.

Various scenarios support different values for $m$. According to Eq.\,\ref{eq:mdot}, the mass entrainment rate $\dot{M}_{\rm{ent}}$ is proportional to $v_{\rm c}^3$. If $\alpha_{\rm{ov}}$ in the step overshoot case most closely corresponds to $\dot{M}_{\rm{ent}}$ in entrainment models, then $m=3$ is appropriate. However, $m=2$ would be supported if $\alpha_{\rm{ov}}$ corresponds best to the penetrability of the boundary ($Ri_{\rm B}$ is inversely proportional to $v_{\rm c}^2$ if $n=1$). The $m=1$ case of $\alpha_{\rm{ov}} \propto v_{\rm c}$ would be similar to the findings of \citet{denissenkov2019}, who reported that the exp-D $f$ parameter scales linearly with the cube root of the convective driving luminosity, or equivalently $f \propto v_{\rm c}$.

\begin{figure}
 \includegraphics[width=\columnwidth]{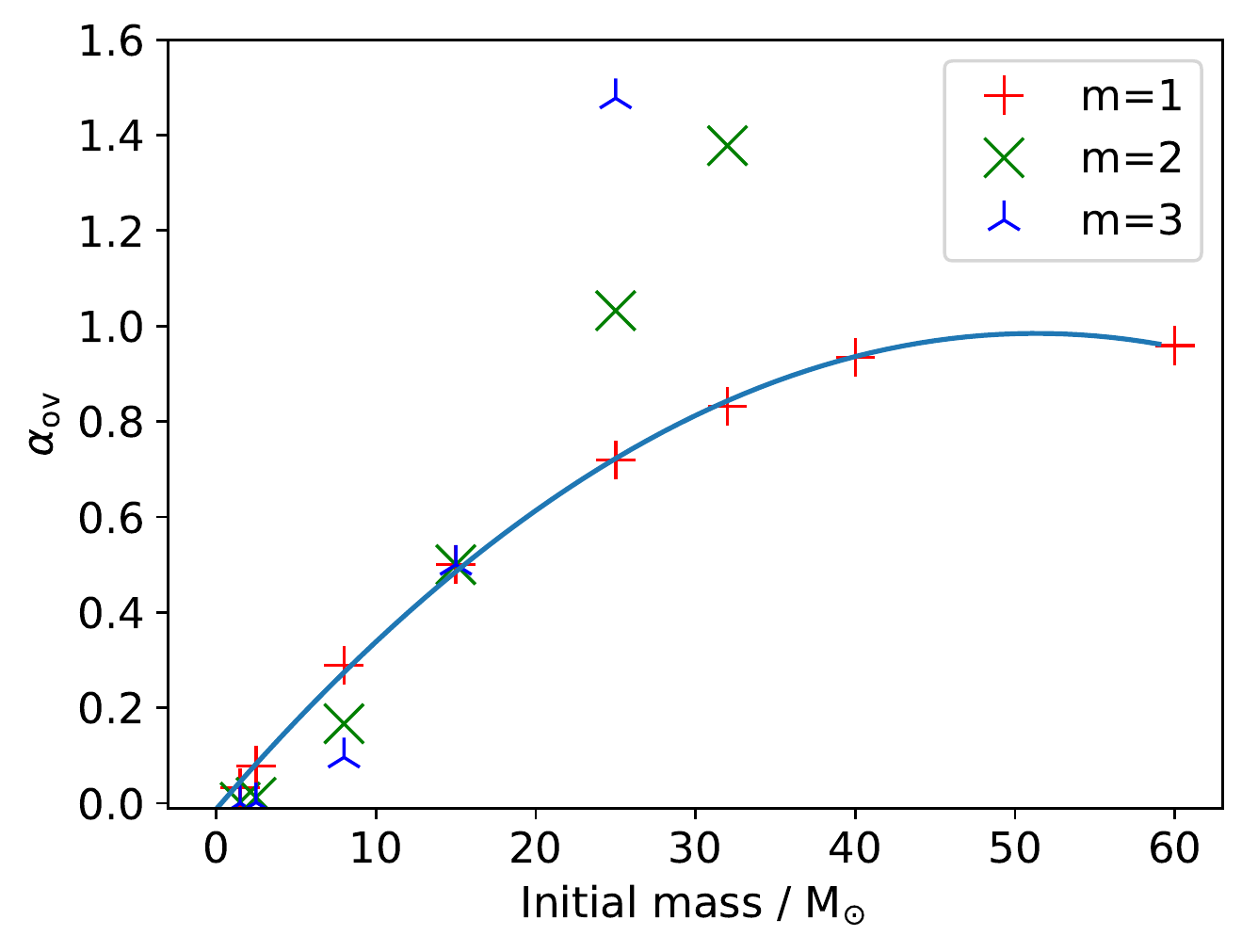}
 \caption{Step overshoot parameter $\alpha_{\rm ov}$ scaled using Eq.\ref{eq:scaledov}. The polynomial fit to the $m=1$ points uses the equation $\alpha_{\rm ov}(M_{\rm{ini}})=-0.00037867 M_{\rm{ini}}^2 +0.03885918 M_{\rm{ini}} -0.01237484$. Previous studies such as \citet{claret2017} and \citet{moravveji2016} show that similar results are obtained using an exp-D $f$ parameter which is roughly a factor of 10 to 15 smaller than the equivalent step overshoot $\alpha_{\rm ov}$. Therefore a fit for $f(M_{\rm{ini}})$ would be roughly 1/10 to 1/15 of $\alpha_{\rm ov}(M_{\rm{ini}})$.}
 \label{fig:scaledov}
\end{figure}

Figure\,\ref{fig:scaledov} shows the predicted values of $\alpha_{\rm{ov}}$ according to Eq.\,\ref{eq:scaledov} with $m=1$, 2 and 3. The $m=2$ and $m=3$ cases quickly reach very high values of $\alpha_{\rm ov}$ above 15\,M$_\odot$; thus the y-axis scale is zoomed onto the lower $\alpha_{\rm ov}$ range. The values of $\alpha_{\rm ov}=0.05$ and $\alpha_{\rm ov}=0.1$ for 1.5\,M$_\odot$ and 2.5\,M$_\odot$ respectively have already been calibrated \citep{ekstrom2012}, but are also underestimated by the $m=2$ and $m=3$ cases. Only $m=1$ matches the already-known values for the lower mass range and does not produce extremely high values in the higher mass range.

Since the $m=1$ case seems the most reasonable, we have provided a polynomial fit to this scaling, described in the caption of Fig.\,\ref{fig:scaledov}. We emphasise that this scaling should only be considered a temporary fix to the problem of mass-dependent CBM and behaviour of the mass range above 60\,M$_{\odot}$ is unknown. Whilst the $m=1$ scaling supports previous findings \citep{denissenkov2019}, the step overshoot values at $M_{\rm{ini}} \geq 30$\,M$_\odot$ are already much larger than the value of $\alpha_{\rm{ov}}=0.5$ favoured by \citet{higgins2019}. Eq.\,\ref{eq:scaledov} also does not take the stiffness of the boundary into account. This may be less of a problem for MS cores, but convective shells which have two boundaries are known to have different stiffnesses for each and different entrainment rates according to the entrainment law \citep{cristini2019}. In addition, the possible mass dependence of $H_{P,\rm{b}}$ (which is used to determine the total overshooting distance, $d_{\rm{ov}}=\alpha_{\rm{ov}} H_{P,\rm{b}}$) should not be discounted, as it also contains information on the stellar structure near the boundary.

Nevertheless, we have calculated an additional model at 25\,M$_{\odot}$ with $\alpha_{\rm{ov}}=0.7$, which is approximately the value suggested by the $m=1$ case of Eq.\,\ref{eq:scaledov}. This model can be found in Table\,\ref{tab:models} and in Figures\,\ref{fig:minteff}, \ref{fig:MHe} and \ref{fig:tauMS} represented by a black clover symbol. Fig.\,\ref{fig:minteff} in particular shows that this model produces a very broad MS with a minimum $\lg{T_{\rm{eff}}}\sim 4$, which is consistent with \citet{castro2014}.

The focus of this study is on entrainment at the convective core boundary during the MS, but many 3D simulations which resulted in entrainment were concerned with later evolutionary phases. The effects of entrainment in post-MS 1D models are unknown, but may be similar to that of other CBM with phenomena such as increased likelihood of convective shell mergers. In convective envelopes, the length-scales and pressure stratification can be significantly different to convective cores. The relatively high importance of thermal diffusion may mean that entrainment is not a suitable CBM prescription in convective envelopes \citep{viallet2015}.

Since our entrainment implementation is cumulative, it is interesting to compare our results to those of \citet{staritsin2013}, who used instantaneous entrainment. Staritsin's values for $A$ were also much smaller than the results of 3D simulations, with $A=4.425\e{-4}$ for the 16\,M$_\odot$ model and $A=4.054\e{-4}$ for the 24\,M$_{\odot}$. This is not dissimilar to our value of $2\e{-4}$, perhaps due to the similarity in calibration: Staritsin required that the entrained distance at the ZAMS was $0.1\,H_P$, guided by asteroseismic results for the star HD\,46202 \citep{briquet2011}. We also based our initial estimate of $A=10^{-4}$ on the MS lifetime of models with $\alpha_{\rm{ov}}=0.1$, as explained in Section\,\ref{sec:parameters}.

However, there are also significant differences between our models and the models of \citet{staritsin2013}. The key result of \citet{staritsin2013} was an entrainment region which decreased with time as the model evolved; we instead see the opposite, since the mass of our entrained region can only ever increase (by construction). As such, Staritsin's entrainment models produced less Helium overall than standard $\alpha_{\rm{ov}}=0.1$ models, whereas our estimations for Helium core sizes were much greater in the entrainment models (see Table\,\ref{tab:models}). In addition, the buoyancy jump continuously increases in \citet{staritsin2013}, whereas we see a plateau in the buoyancy jump near the middle of the MS (as explained in Section\,\ref{sec:time_dependence}. This difference could be due to the buoyancy jump integration distance used by Staritsin, $h \sim 2v_{\rm c}/N$. Since $v_{\rm c}$ grows with time during the MS (in Staritsin's models as well as ours), the integration length $h$ would similarly increase with time, potentially leading to the increase in $\rm{\Delta}b$.

To conclude, the entrainment law, through its dependence on the bulk Richardson number, produces models with a wider MS for high mass stars than standard models. In addition, the extension of the MS increases with mass, as required by observation. However, the value of the entrainment law $A$ parameter required to produce the correct MS width for the lower mass stars in our grid is orders of magnitude smaller than the value derived from 3D simulations of convection in the later stages of stellar evolution. This value may change further if more aspects of convective boundary physics are included, such as shear.
Although these models are not complete, they are an important step in the right direction since they show that convective boundary penetrability is a key part of the physics behind the mass dependence of CBM.


 

\section*{Acknowledgements}


The authors acknowledge support from EU FP7 ERC 2012 St
Grant 306901. R.H. acknowledges support from the World
Premier International Research Centre Initiative (WPI Initiative),
MEXT, Japan and the IReNA AccelNet
Network of Networks, supported by the National Science Foundation under Grant
No. OISE-1927130. This article is based upon work from the
ChETEC COST Action (CA16117), supported by COST
(European Cooperation in Science and Technology). C.G.,
R.H., C.M. and S.E. thank ISSI, Bern, for their support on organizing
meetings related to the content of this paper. C.G. and S.E. acknowledge the STAREX grant from the ERC Horizon 2020 research and innovation programme (grant agreement No 833925). This work used the DiRAC@Durham facility managed by
the Institute for Computational Cosmology on behalf of the
STFC DiRAC HPC Facility (www.dirac.ac.uk). The equipment
was funded by BEIS capital funding via STFC capital grants
ST/P002293/1 and ST/R002371/1, Durham University, and
STFC operations grant ST/R000832/1. This work also used
the DiRAC Data Centric system at Durham University,
operated by the Institute for Computational Cosmology on
behalf of the STFC DiRAC HPC Facility. This equipment was
funded by BIS National E Infrastructure capital grant ST/
K00042X/1, STFC capital grants ST/H008519/1 and ST/
K00087X/1, STFC DiRAC Operations grant ST/K003267/1,
and Durham University. DiRAC is part of the National E
Infrastructure.

\section*{Data Availability}
The data underlying this article will be shared on reasonable request to the corresponding author.




\bibliographystyle{mnras}
\bibliography{biblio} 




\appendix

\section{Model Resolution}
\label{app}
\subsection{Spatial Resolution}
Spatial resolution in the Geneva code is set by parameters controlling the allowed change in variables between model grid points.
The controlled variables include pressure, luminosity, and the chemical species $^4$He, $^{16}$O, and $^{20}$Ne.
If a variable $q$ is controlled by the resolution parameter \texttt{dgrq}, an extra grid point is added between points $i$ and $i+1$ when the following condition is met:
\begin{equation}
\label{eq:dgrq}
|q_i-q_{i+1}|>\texttt{dgrq}.
\end{equation}
This results in the addition of grid points where the variable $q$ changes quickly.

The convergence of the MS lifetime with adjustment of the resolution parameters was used to judge good resolution.
MS lifetime was chosen due to its relationship with MS width in the Hertzsprung-Russell diagram and its sensitivity to core size.
The $^4$He abundance was therefore identified as the most important variable for spatial resolution, having the greatest impact on number of grid points in the region of interest (the core boundary) compared to the other controlled variables.
Other resolution parameters had little effect on MS lifetime.

\begin{figure}
        \includegraphics[width=\columnwidth]{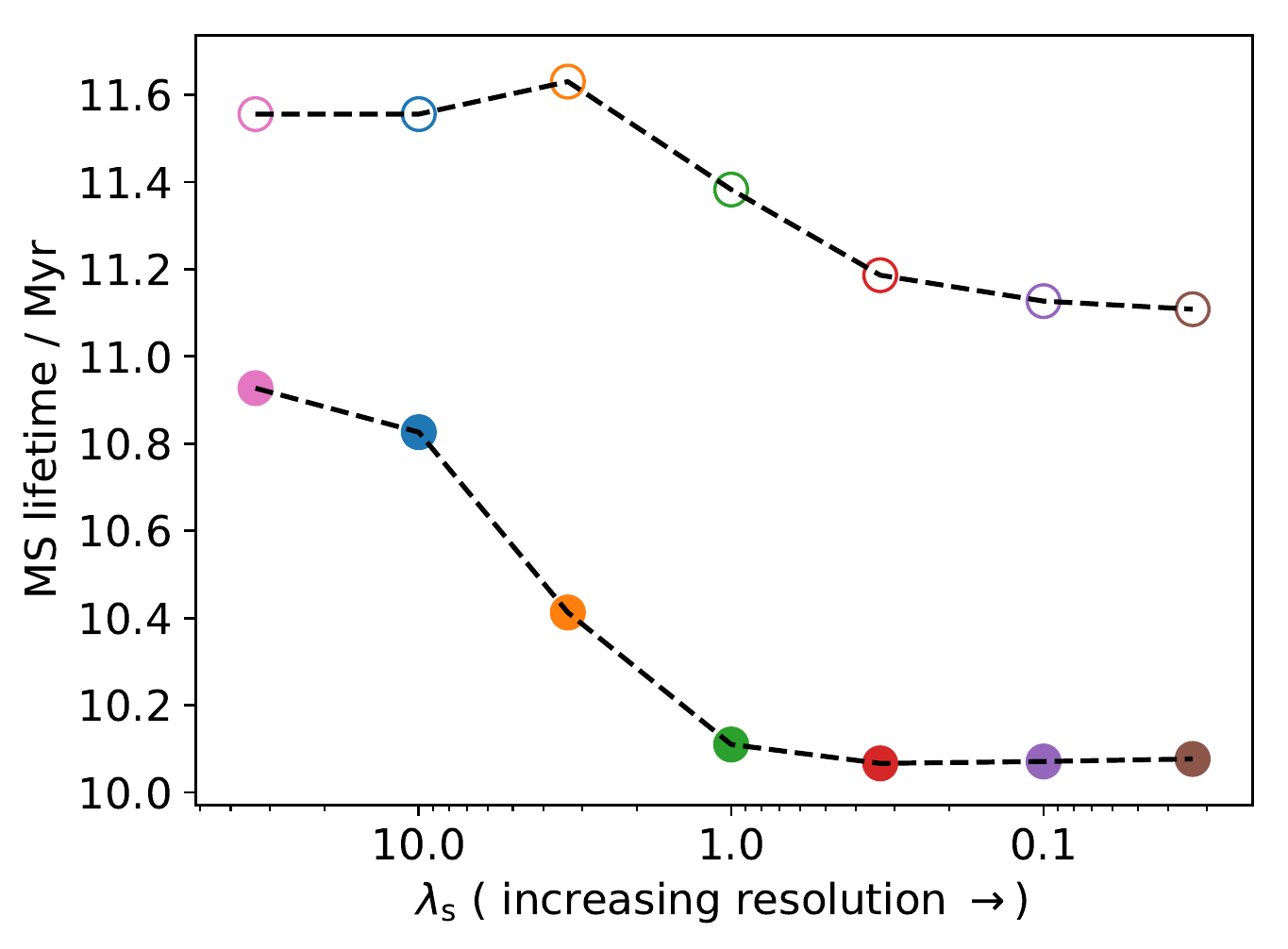}
    \caption{MS lifetime of 15\,M$_{\odot}$ models with varying spatial resolution. The filled circles represent step overshoot models with $\alpha_{\rm{ov}}=0.1$ whereas the open circles represent entrainment models with $A=10^{-4}$ and $n=1$.}
    \label{fig:spatial_res}
\end{figure}

We set the $^4$He resolution parameter $\texttt{dgry}=0.003$ for all models in our grid. This was based on the lifetimes of the step overshoot models shown with filled circles in Figure\,\ref{fig:spatial_res}.
This figure shows how MS lifetime varies for a 15\,M$_{\odot}$ model when changing $\texttt{dgry}=0.003$ by a factor $\lambda_{\mathrm{s}}$ (note that these MS lifetimes were calculated using models with $\lambda_{\rm t}=3$; see Section\,\ref{sec:time_res}). Although the resolution was chosen based on the convergence of the step overshoot models before calculating any entrainment models, a similar set of lifetimes for the entrainment models is shown in Fig.\,\ref{fig:spatial_res} with open circles for comparison. Table\,\ref{tab:spatial} gives the mean number of grid points in the models presented in Fig.\,\ref{fig:spatial_res}.

Our use of integrals of the buoyancy jump in the calculation of $Ri_{\mathrm{B}}$ require particularly fine resolution at the convective boundary to ensure that $Ri_{\mathrm{B}}$ remains as stable as possible.
Otherwise, $Ri_{\mathrm{B}}$ can increase sharply over a single time step before dropping again.
Whilst this seems to be a transient effect that does not impact the MS lifetime, it can make analysis of the behaviour of $Ri_{\mathrm{B}}$ difficult.
We therefore use another resolution condition which adds a layer if
\begin{equation}
 \label{eq:dgrra}
 |\ln r_{i} - \ln r_{i+1}|>\texttt{dgrra}
\end{equation}
within a small mass region centred on the furthest extent of CBM.
We use 1\% of the total mass as the size of this region, which is both large enough to accommodate changes in the position of the boundary as the model converges and small enough to not impact MS lifetime.
We use $\texttt{dgrra}=0.0003$ as this generally produces the best-behaved $Ri_{\mathrm{B}}$.

\begin{table}
 \centering
 \caption{Mean number of grid points over the main sequence for the step overshoot and entrainment models presented in Fig.\,\ref{fig:spatial_res}.}
 \label{tab:spatial}
 \begin{tabular}{ccc}
  \hline
  $\lambda_s$ & \multicolumn{2}{c}{number of grid points}\\
  & step overshoot & entrainment\\
  \hline
  $100/3$ & 199 & 273 \\
  10 & 206 & 273 \\
  10/3 & 231 & 272 \\
  1 & 380 & 384 \\
  1/3 & 577 & 874 \\
  1/10 & 797 & 831 \\
  1/30 & 715 & 883 \\
  \hline
 \end{tabular}
\end{table}

\subsection{Time Resolution}
\label{sec:time_res}
\begin{figure}
        \includegraphics[width=\columnwidth]{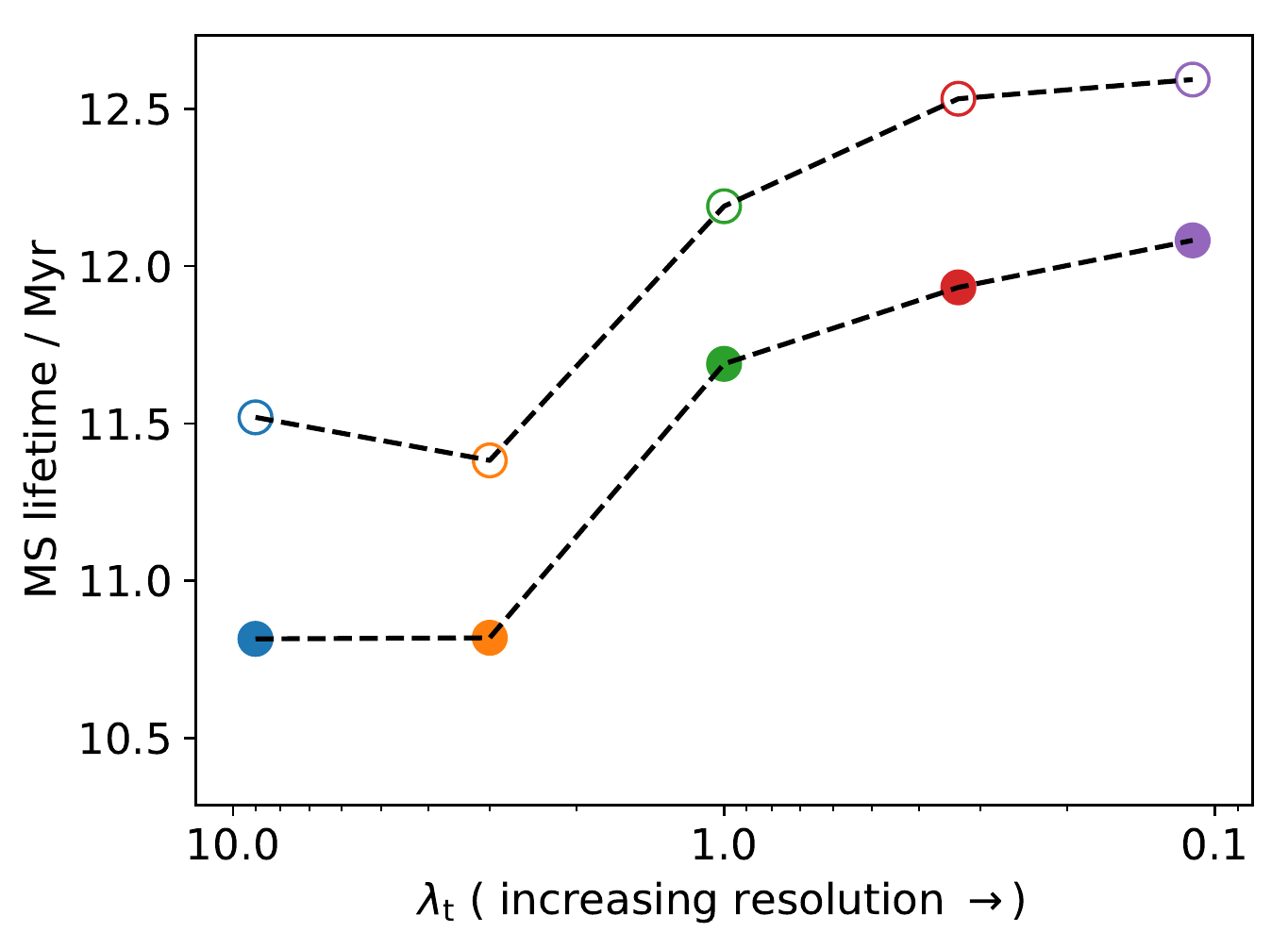}
    \caption{MS lifetime of 15\,M$_{\odot}$ models with varying temporal resolution. The filled circles represent step overshoot models with $\alpha_{\rm{ov}}=0.1$ whereas the open circles represent entrainment models with $A=10^{-4}$ and $n=1$.
}
    \label{fig:temporal_res}
\end{figure}

The time step in \texttt{GENEC} is controlled using the energy generation rate in the centre. It is generally set so that the MS is split in several thousand time steps. 
Figure \ref{fig:temporal_res} shows the effect on MS lifetime of changing time step length by a factor $\lambda_{\mathrm{t}}$ in a spatially resolved ($\lambda_{\mathrm{s}}=1$) 15\,M$_{\odot}$ model. The total number of time steps for each of these models is given in Table\,\ref{tab:temporal}. As with the spatial resolution, we chose a temporal resolution based on the lifetimes of the step overshoot models shown in Fig.\,\ref{fig:temporal_res} before calculating our grid. However, the entrainment model lifetimes shown for comparison display a similar behaviour to the step overshoot models.
The temporal resolution corresponding to $\lambda_{\mathrm{t}}=1$ results in $\sim10^4$ time steps in most cases, although the 1.5\,M$_{\odot}$ models generally have around half the number of steps compared to the other masses.

\begin{table}
 \centering
 \caption{Total number of time steps over the main sequence for the step overshoot and entrainment models presented in Fig.\,\ref{fig:temporal_res}.}
 \label{tab:temporal}
 \begin{tabular}{ccc}
  \hline
  $\lambda_t$ & \multicolumn{2}{c}{number of time steps}\\
  & step overshoot & entrainment\\
  \hline
  9 & 1292 & 881 \\
  3 & 3581 & 3855  \\
  1 & 10246 & 10913 \\
  1/3 & 30302 & 36155 \\
  1/9 & 91154 & 90261 \\
  \hline
 \end{tabular}
\end{table}


\bsp	
\label{lastpage}
\end{document}